\documentclass[twocolumn,pra]{revtex4}%
\usepackage[pdftex,colorlinks=false,bookmarks]{hyperref}
\usepackage{amsfonts}
\usepackage{amsmath}
\usepackage{amssymb}
\usepackage{graphicx}
\usepackage{color}%
\providecommand{\U}[1]{\protect\rule{.1in}{.1in}}

\ifx\pdfoutput\relax\let\pdfoutput=\undefined\fi
\newcount\msipdfoutput
\ifx\pdfoutput\undefined\else
\ifcase\pdfoutput\else
\ifx\paperwidth\undefined\else
\ifdim\paperheight=0pt\relax\else\pdfpageheight\paperheight\fi
\ifdim\paperwidth=0pt\relax\else\pdfpagewidth\paperwidth\fi
\fi\fi\fi
\begin{document}
\title{Distributed Quantum Computation Based-on Small Quantum Registers}
\author{Liang Jiang$^{1}$, Jacob M. Taylor$^{1,2}$, Anders S. S\o rensen$^{3}$,
Mikhail D. Lukin$^{1}$}
\affiliation{$^{1}$ Department of Physics, Harvard University, Cambridge, Massachusetts 02138}
\affiliation{$^{2}$ Department of Physics, Massachusetts Institute of Technology,
Cambridge, Massachusetts 02139}
\affiliation{$^{3}$ Quantop and The Niels Bohr Institute, University of Copenhagen, DK-2100
Copenhagen \O , Denmark}
\date{\today}


\begin{abstract}
We describe and analyze an efficient register-based hybrid quantum computation
scheme. Our scheme is based on probabilistic, heralded optical connection
among local five-qubit quantum registers. We assume high fidelity local
unitary operations within each register, but the error probability for
initialization, measurement, and entanglement generation can be very high
($\sim5\%$). We demonstrate that with a reasonable time overhead our scheme
can achieve deterministic non-local coupling gates between arbitrary two
registers with very high fidelity, limited only by the imperfections from the
local unitary operation. We estimate the clock cycle and the effective error
probability for implementation of quantum registers with ion-traps or
nitrogen-vacancy (NV) centers. Our new scheme capitalizes on a new efficient
two-level pumping scheme that in principle can create Bell pairs with
arbitrarily high fidelity. We introduce a Markov chain model to study the
stochastic process of entanglement pumping and map it to a deterministic
process. Finally we discuss requirements for achieving fault-tolerant
operation with our register-based hybrid scheme, and also present an
alternative approach to fault-tolerant preparation of GHZ states.

\end{abstract}
\maketitle


\section{Introduction}

The key challenge in experimental quantum information science is to identify
isolated quantum mechanical systems with good coherence properties that can be
manipulated and coupled together in a scalable fashion. Recently, considerable
advances have been made towards interfacing of individual qubits in the
optical and microwave regimes. These include advances in cavity QED
\cite{Mabuchi02,Wallraff04} as well as in probabilistic techniques for
entangling remote qubits \cite{Simon03,Duan04,Lim06}. At the same time,
substantial progress has been made towards the physical implementation of
few-qubit quantum registers using systems of coupled trapped
ions~\cite{Barrett04,Riebe04,Reichle06}, neutral atoms \cite{Mandel03}, or
solid-state qubits based on either electronic and nuclear spins in
semiconductors~\cite{Jelezko04,Childress06,Dutt07} or superconducting islands
\cite{Yamamoto03,McDermott05}.

\begin{figure}[tb]
\begin{center}
\includegraphics[
width=8 cm
]{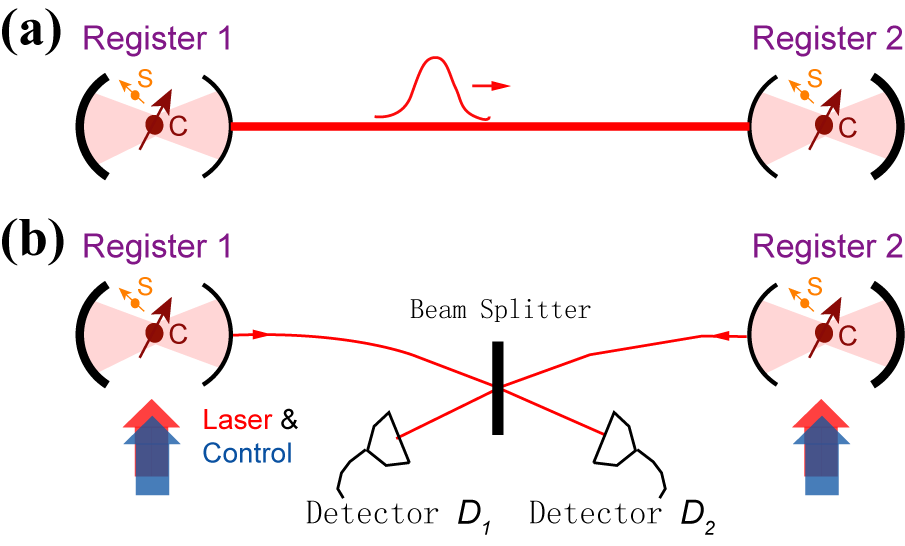}
\end{center}
\caption[fig:ENG]{(Color online) Two schemes to couple different registers.
(a) Deterministic state transfer from one register to the other \cite{Cirac97}%
. (b) Probabilistic entanglement generation, heralded by distinct detector
click patterns \cite{Childress05,Duan03,Simon03}. In the text, we argue that
probabilistic, heralded entanglement generation is sufficient for
deterministic distributed quantum computation.}%
\label{fig:ENG}%
\end{figure}

While the precise manipulation of large, multi-qubit systems still remains an
outstanding challenge, various approaches for connecting such few qubit
sub-systems into large scale circuits have been investigated
\cite{Dur03,Lim06,Oi06,VanMeter06,VanMeter07}. These studies suggest that
hybrid schemes, which benefits from short range interactions for local
coupling and (optical) long range interactions for non-local coupling, might
be an effective way toward large scale quantum computation: small local
few-qubit quantum systems may be controlled with very high precession using
optimal control techniques \cite{Leibfried03,Vandersypen04}, and in practice
it may be more feasible to operate several such small sub-systems compared to
the daunting task of high-precession control of a single large quantum system
with thousands of qubits. Optical techniques for quantum communication can
then be used to connect any two sub-systems. For example, we may directly
transfer a quantum state from one sub-system to another via an optical channel
\cite{Cirac97} (see Fig.~\ref{fig:ENG}a), which immediately provides an
efficient way to scale-up the total number of the physical qubits we can
manipulate coherently. In particular the use of optical means for connecting
different subsystems has the advantage that it allows for fast non-local
operations over large distances. This is advantageous for quantum error
correction since the existence of such non-local coupling operations
alleviates the threshold requirement for fault-tolerant quantum computation
\cite{Svore05}.

In practice, however, it is very difficult to have a perfect optical
connection. In particular, there is excitation loss associated with the
optical channel, due to scattering or absorption. For a lossy channel, it is
therefore more desirable to use it to generate entanglement between different
sub-systems (see Fig.~\ref{fig:ENG}b), rather than for direct state transfer.
The entanglement generation is then heralded by the click patterns from the
photon detectors. Such detection-based scheme is intrinsically robust against
excitation loss in the channel, since it only reduces the success probability
but does not affect the entanglement fidelity. This entanglement can then be
used as a resource to teleport quantum state from one sub-system to another
\cite{Bouwmeester00}. More generally, entanglement provides a physical
resource to implement non-local unitary coupling gates (such as CNOT gate)
\cite{Gottesman99,Sorensen98,Zhou00,Eisert00}. If there is only one physical
qubit for each sub-system, so-called cluster states \cite{Briegel00} can be
created based on probabilistic, heralded entanglement generation (see Ref.
\cite{Lim06} and references therein). Such cluster states can be used for
universal quantum computation \cite{Raussendorf00}. If there are two physical
qubits available for each sub-system, cluster states can be created
deterministically \cite{Benjamin06}; meanwhile one can also use these
two-qubit sub-systems to implement any quantum circuit directly \cite{Duan04}.

Furthermore, realistic optical channel connecting sub-systems has other
imperfections beside excitation loss, such as the distortion of the
polarization or shape of the wave-packet. These imperfections will reduce the
fidelity of the heralded entanglement generated between the sub-systems. To
overcome these imperfections in the channel, entanglement purification schemes
have been proposed, which may create some high-fidelity Bell pairs from many
low-fidelity ones \cite{Bennett96a,Deutsch96}. In particular, the entanglement
pumping scheme originally presented in the context of quantum communication
over long distances \cite{Briegel98,Dur99,Childress05} provides a very
efficient purification scheme in terms of local physical resources, and
D{\"{u}}r and Briegel \cite{Dur03} first proposed to use such entanglement
pumping scheme for quantum computation. In principle, the infidelity of the
purified Bell pair shared by the sub-systems can be very low, and is only
limited by the error probability from local operations. In Ref. \cite{Dur03}
it was found that \emph{three} auxiliary qubits (requiring a total of five
qubits including the storage and communication qubits) for each sub-system
provide enough physical resources to obtain high fidelity entangled pairs via
entanglement pumping.

In order to implement the idea of distributed quantum computation using
realistic optical channel and imperfect operations, it is necessary to
consider the following questions: What are the minimal local physical
resources needed for robust entanglement generation? What is the time overhead
associated with entanglement generation? Can we extend the robustness to other
imperfections, such as errors from initialization and measurement?

Motivated by these considerations, we study the practical implementation of
robust quantum registers for scalable applications. In Ref. \cite{JTSL07} we
have proposed an entanglement purification scheme that only requires
\emph{two} auxiliary qubits for robust entanglement generation. We have found
that the time overhead associated with entanglement generation ranges from a
factor of $10$ to a few $100$, depending on the initial and targeting
infidelities. We have also suggested to use \emph{one more} auxiliary qubit to
suppress errors from initialization and measurement. Thus, our hybrid scheme
also requires only 5 (or fewer) qubits with local deterministic coupling,
while providing additional improvements over the protocol of Ref.
\cite{Dur03}: reduced measurement errors, higher fidelity, and more efficient
entanglement purification. In this paper, we will provide a detailed
discussion on the register-based, hybrid quantum computation scheme presented
in Ref. \cite{JTSL07}.

The paper is organized as follows. In Sec.
\ref{Quantum Register and Experimental Implementations}, we will introduce the
concept of quantum register and discuss two experimental implementations. In
Sec. \ref{Two-Qubit Register}, we will review the idea of universal quantum
computation based on two-qubit quantum registers. In Sec.
\ref{Errors and Imperfections}, we will specify the error models for
imperfections and provide the basic ideas underlying our robust operations. In
Sec. \ref{Robust Measurement & Initialization}, we will describe the robust
measurement/initialization scheme. In Sec.
\ref{Robust Non-Local Two-Qubit Gate}, we will present our bit-phase two-level
entanglement pumping scheme. In Sec. \ref{Markov Chain Model}, we will
introduce the Markov chain model to quantitatively analyze the time overhead
and residual infidelity associated with the stochastic process of entanglement
pumping, and discuss further improvement upon our two-level entanglement
pumping scheme. In Sec. \ref{Mapping to Deterministic Model}, we will map our
stochastic, hybrid, and distributed quantum computation scheme to a
deterministic computation model that is characterized by two quantities (the
clock cycle and effective error probability), and estimate the practical
values for these quantities. We will also consider the constraint set by the
finite memory lifetime and determine the achievable performance of hybrid
distributed quantum computation. Finally, in Sec. \ref{Fault Tolerance}, we
will discuss using our hybrid scheme for fault-tolerant quantum computation
with quantum error correcting codes, and provide a resource-efficient approach
for fault-tolerant preparation of the GHZ states.

\section{Quantum Register and Experimental
Implementations\label{Quantum Register and Experimental Implementations}}

\begin{figure}[tb]
\begin{center}
\includegraphics[
width=8 cm
]{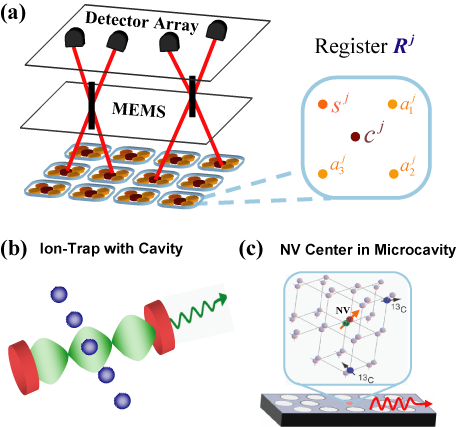}
\end{center}
\caption[fig:DistributedQC]{(Color online) Distributed quantum computer. (a)
Illustration of distributed quantum computer based on many quantum registers.
Each register has five physical qubits, including one communication qubit
($c$), one storage qubit ($s$), and three auxiliary qubits ($a_{1,2,3}$).
Local operations for qubits within the same register have high fidelity.
Entanglement between non-local registers can be generated probabilistically
\cite{Childress05,Duan03,Simon03}. Devices of optical micro-eletro-mechanical
systems (MEMS) \cite{Kim07} can efficiently route photons and couple arbitrary
pair of registers. Detector array can simultaneously generate entanglement for
many pairs of registers. (b) An ion trap coupled to a cavity also provides a
promising candidate for distributed quantum computation. A single ion is
resonantly coupled to the cavity and serves as the communication qubit; while
the others can be storage or auxiliary qubits. (c) Nitrogen-vacancy (NV)
defect center in photonic crystal micro-cavity. The inset shows the atomic
structure of the NV center \cite{Jelezko04}, which forms a quantum register.
The electronic spin localized at the vacancy is optically accessible
(measurement/initialization) and can play the role of the communication qubit.
The nuclear spins from proximal $^{13}$C atoms constitute the storage and
auxiliary qubits, which are coherently controlled via hyperfine interaction
and rf pulses. }%
\label{fig:DistributedQC}%
\end{figure}

We define a \emph{quantum register} as a few-qubit device (see
Fig.~\ref{fig:DistributedQC}a) that contains one \emph{communication} qubit
($c$), with a photonic interface; one \emph{storage} qubit ($s$), with very
good coherence times; and several \emph{auxiliary} qubits ($a_{1}%
,a_{2},......$), used for purification and error correction. A critical
requirement for a quantum register is high-fidelity unitary operations between
the qubits within a register.

The quantum registers considered here can be implemented in several physical
systems, but in this paper we shall focus on two specific systems where these
considerations are particularly relevant. First, ion traps have been used to
demonstrate all essential elements of quantum registers. (1) The ion qubits
may play the role of communication qubits: they can be initialized and
measured efficiently using optical pumping and cycling transitions
respectively, and they can also be prepared in highly entangled states with
the polarization of single photons \cite{Blinov04,Rosenfeld07}. Very recently,
entanglement generation between ion qubits from two remote traps has been
demonstrated \cite{Maunz07,Moehring07}. This experiment directly demonstrates
the non-local connection required for our hybrid approach. Since the photon
collection and detection efficiency is not perfect, the entanglement
generation is a probabilistic process. However, the entanglement generation is
also a heralded process, because different click patterns from the detectors
can be used to identify each successful entanglement generation. As we will
discuss Sec. \ref{Two-Qubit Register}, such probabilistic, heralded
entanglement generation process is already sufficient to implement
deterministic non-local coupling gates. (2) The ion qubits can be good storage
qubits as well. Coherence time of approximately $10$ seconds has been
demonstrated in ion traps \cite{Langer05,Haffner05b}, which is $10^{6\sim7}$
times longer than the typical gate operation time which is at the order of
$\mu s$. Since fault-tolerant quantum computation only requires the coherence
time to be approximately $10^{4}$ times longer than the gate time, the very
long coherence of the ions provides new opportunities, such as performing
non-local coupling gates with some extra time overhead. (3) Coherent
manipulation of few ions in the same trap has been demonstrated
\cite{Leibfried05,Haffner05}, allowing gates to be implemented among the
qubits in the register. (4) High fidelity operations between the ion qubits
within the ion trap has also been demonstrated \cite{Leibfried03}.

A second promising candidate for implementing quantum registers is the
nitrogen vacancy (NV) centers in diamond. Each NV center can be regarded as an
ion trap confined by the diamond crystal, which can be treated as a single
register. The qubits for each NV register consists of one electronic spin
associated with the defect and several nuclear spins associated with the
proximal C-13 nuclei. The electronic spin is optically active, so that it can
be measured and initialized optically. With optical cavities or diamond-based
photonic crystal micro-cavities \cite{Tomljenovic-Hanic06} one could enhance
photon collection efficiency towards unity. Furthermore, the electron spin can
be coherently manipulated by microwave pulses \cite{Jelezko04}. The electronic
spin is thus suitable as the communication qubit. The nuclear spins are
coupled to the electronic spin via hyperfine interactions. One can either use
these hyperfine interaction to directly rotate the nuclear spins
\cite{Hodges07, Khaneja07}, or apply radio-frequency pulses to address
individual nuclear spins spectroscopically \cite{CJKL}. These nuclear spins
have very long coherence times approaching seconds \cite{Dutt07}, and can be
good storage and auxiliary qubits. Furthermore, the optical manipulation of
the electronic spin can be well decoupled from the nuclear spins \cite{JDL07}.
It can be inferred from the recent experiment \cite{Dutt07} that the fidelity
of local operations between electronic and nuclear spins is higher than
$90\%$. While the fidelity is still low for the procedures considered here, we
believe that it can be significantly improved (to higher than $0.999$) by
optimal control techniques \cite{Vandersypen04,CJKL}, such as composite pulses
\cite{Levitt86} and numerically optimized GRAPE pulses \cite{Khaneja05}.

\section{Universal Quantum Computation with Two-Qubit Registers:
Fundamentals\label{Two-Qubit Register}}


We now consider universal quantum computation via the simplest possible
two-qubit registers \cite{Duan04, Benjamin06}. Each register has one qubit for
communication and the other qubit for storage. We can use probabilistic
approaches from quantum communication (\cite{Childress05} and references
therein) to generate entanglement between communication qubits from two
\emph{arbitrary} non-local registers. The probabilistic entanglement
generation creates a Bell pair conditioned on certain measurement outcomes,
which are distinct from outcomes of unsuccessful entanglement generation. If
the entanglement generation fails, it can be re-attempted until success, with
an exponentially decreasing probability of continued failure.

\begin{figure}[tb]
\begin{center}
\includegraphics[
width=8 cm
]{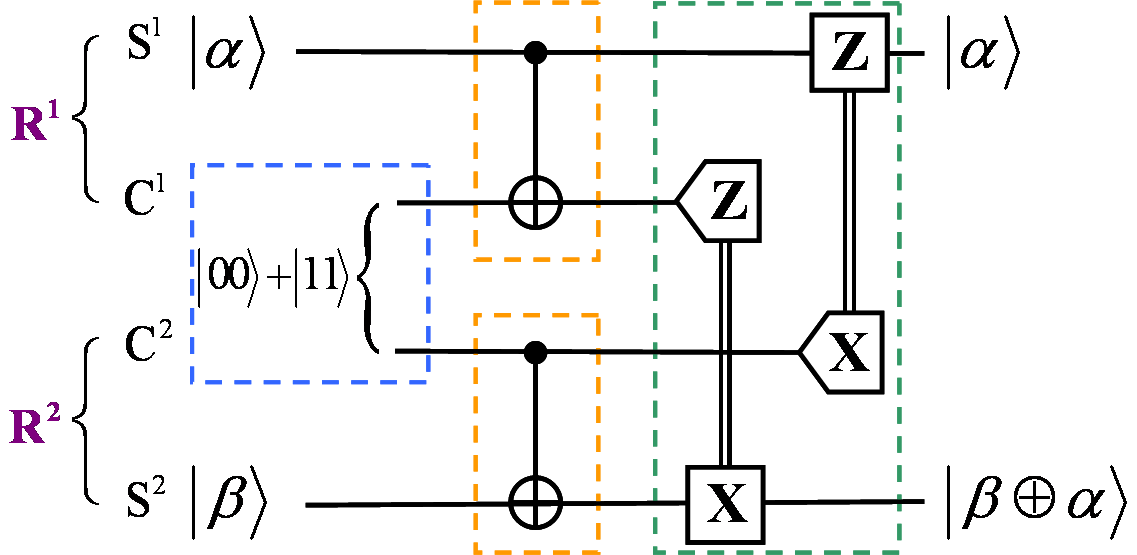}
\end{center}
\caption[fig:CNOT]{(Color online) Quantum circuit for non-local CNOT gate
between two registers $R^{1}$ and $R^{2}$. The circuit starts with the Bell
state $\left\vert \Phi^{+}\right\rangle _{c^{1},c^{2}}=\left(  \left\vert
00\right\rangle +\left\vert 11\right\rangle \right)  /\sqrt{2}$ for the
communication qubits $c^{1}$ and $c^{2}$ (the left blue box). After local
unitary operations within each register (the middle orange boxes), qubits
$c^{1}$ and $c^{2}$ are projectively measured in $Z$ and $X$ basis,
respectively (the right green box). According to Eq.~(\ref{eq:CNOT}), up to
some local unitary gates (the right green box), this circuit implements the
non-local CNOT gate to qubits $s^{1}$ and $s^{2}$ from two quantum registers.}%
\label{fig:CNOT}%
\end{figure}

When the communication qubits ($c^{1}$ and $c^{2}$) are prepared in the Bell
state, we can immediately perform the non-local CNOT gate on the storage
qubits ($s^{1}$ and $s^{2}$) using gate-teleportation between registers
$R^{1}$ and $R^{2}$. The gate-teleportation circuit in Fig.~\ref{fig:CNOT}
implements (before the conditional Pauli operations) the following map
\begin{equation}
\left\vert \phi\right\rangle _{s^{1},s^{2}}\left\vert \Phi^{+}\right\rangle
_{c^{1},c^{2}}\rightarrow(\sigma_{s^{1}}^{z})^{m_{c^{2}}}(\sigma_{s^{2}}%
^{x})^{m_{c^{1}}}\mathrm{CNOT}_{s^{1},s^{2}}\left\vert \phi\right\rangle
_{s^{1},s^{2}}, \label{eq:CNOT}%
\end{equation}
where $\left\vert \Phi^{+}\right\rangle _{c^{1},c^{2}}=\left(  \left\vert
00\right\rangle +\left\vert 11\right\rangle \right)  /\sqrt{2}$,
$\mathrm{CNOT}_{i,j}$ is the controlled-NOT (CNOT) gate with the $i$th qubit
as control and the $j$th qubit as target, and $m_{i}=0,1$ is the measurement
result for qubit $i$ from the circuit in Fig.~\ref{fig:CNOT}.

\begin{figure}[tb]
\begin{center}
\includegraphics[
width=8 cm
]{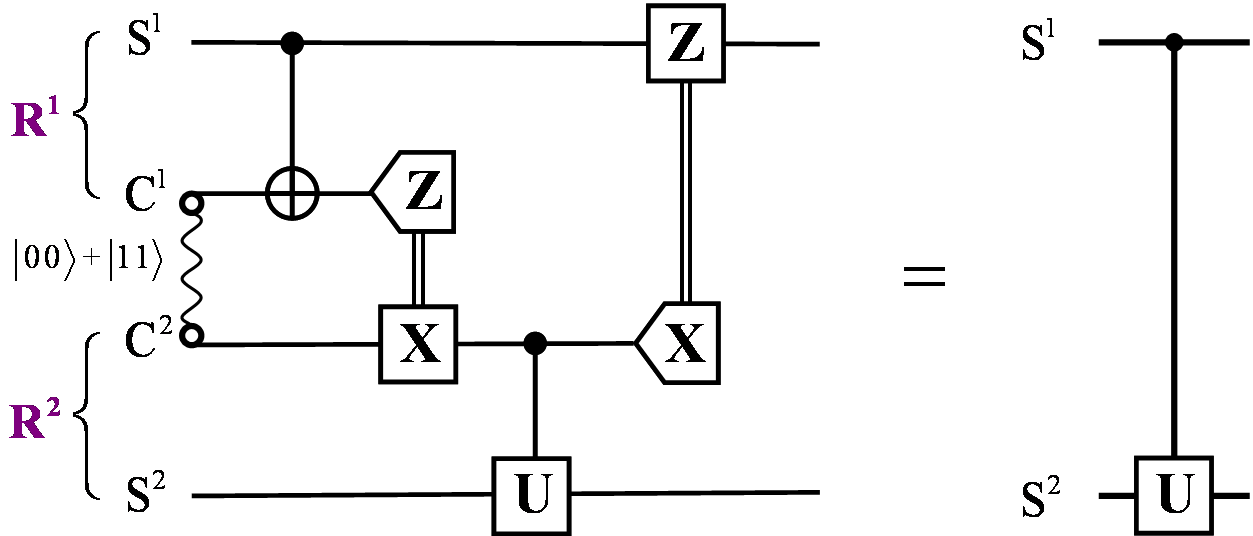}\label{fig:controlled-U}
\end{center}
\caption[fig:Controlled-U]{(Color online) General circuit for non-local
Controlled-U gate \cite{Eisert00}, with the storage qubit $S^{1}$ as control
and the storage qubit $S^{2}$ as target. One Bell pair is consumed for this
operation.}%
\label{fig:Controlled-U}%
\end{figure}

By consuming one Bell pair, one can implement any non-local controlled-U gate
between two storage qubits \cite{Eisert00}, as shown in
Fig.~\ref{fig:Controlled-U}. Since operations on a single qubit can be
performed within a register, the CNOT operation between different quantum
registers is in principle sufficient for universal quantum computation
\cite{Duan04}. Similar approaches are also known for deterministic generation
of graph states \cite{Benjamin06} ---an essential resource for one-way quantum
computation \cite{Raussendorf00}.

We emphasize that deterministic entanglement generation is \emph{not}
required, which opens up a wide range of possibilities of entanglement
generation. It is experimentally challenging to achieve deterministic quantum
state transfer directly \cite{Cirac97}, but we are able to achieve the same
task by probabilistic entanglement scheme and two-qubit quantum registers
\cite{Duan04}.

\section{Errors and Imperfections\label{Errors and Imperfections}}

In practice, the qubit measurement, initialization, and entanglement
generation can be noisy with error probabilities as high as $\sim5\%$, due to
practical limitations such as imperfect cycling transitions, finite collection
efficiency, and poor interferometric stability. As a result, there will be a
large error probability in non-local gate circuits. In contrast, local unitary
operations may fail infrequently ($p_{L}\lesssim10^{-4}$) when quantum control
techniques for small quantum system are utilized
\cite{Vandersypen04,Leibfried03}. We now show that the most important sources
of imperfections, such as imperfect initialization and measurement errors for
individual qubits in each quantum register, and entanglement generation errors
between registers, can be corrected with a modest increase in register size.
We determine that with just \emph{three} additional auxiliary qubits and
high-fidelity local unitary operations, all these errors can be efficiently
suppressed by repeated quantum non-demolition (QND) measurement
\cite{Braginsky80} and entanglement purification \cite{Briegel98,Dur99}. This
provides an extension of Ref. \cite{Dur03} that mostly focused on suppressing
errors from entanglement generation.

We will use the following error model for the entire paper: (1) The imperfect
local two-qubit operation $U_{ij}$ is
\begin{equation}
U_{ij}\rho U_{ij}^{\dag}\rightarrow\left(  1-p_{L}\right)  U_{ij}\rho
U_{ij}^{\dag}+\frac{p_{L}}{4}\operatorname*{Tr}\nolimits_{ij}\left[
\rho\right]  \otimes\mathbf{I}_{ij} \label{eq:ErrorModel1}%
\end{equation}
where $\operatorname*{Tr}_{ij}\left[  \rho\right]  $ is the partial trace over
the qubits $i$ and $j$, and $\mathbf{I}_{ij}$ is the identity operator for
qubits $i$ and $j$. This error model describes that with a probability
$1-p_{L}$ the gates perform the correct operation and with a probability
$p_{L}$ the gates produce a complete random output for the two involved qubits
\footnote[20]{The error model introduced in
Eq.~(\ref{eq:ErrorModel1}) can be regarded as the worst case error, since it
in principle includes all possible errors that can happen to the
system.}. (2) The imperfect initialization of state $\left\vert
0\right\rangle $ will prepare a mixed state
\begin{equation}
\rho_{0}=\left(  1-p_{I}\right)  \left\vert 0\right\rangle \left\langle
0\right\vert +p_{I}\left\vert 1\right\rangle \left\langle 1\right\vert ,
\label{eq:ErrorModel2}%
\end{equation}
which has error probability $p_{I}$, i.e., it prepares the wrong state with a
probability $p_{I}$. (3) The imperfect measurement of state $\left\vert
0\right\rangle $ will correspond to the projection operator
\begin{equation}
P_{0}=\left(  1-p_{M}\right)  \left\vert 0\right\rangle \left\langle
0\right\vert +p_{M}\left\vert 1\right\rangle \left\langle 1\right\vert ,
\label{eq:ErrorModel3}%
\end{equation}
This operator describes that a qubit prepared in state $|0\rangle$ or
$|1\rangle$ will give rise to the opposite measurement output with the
measurement error probability $p_{M}$. (4) Finally, the entanglement fidelity
for a non-ideal preparation of state $\left\vert \Phi^{+}\right\rangle $ is
defined as
\begin{equation}
F=\left\langle \Phi^{+}\right\vert \rho\left\vert \Phi^{+}\right\rangle ,
\label{eq:DefFidelity}%
\end{equation}
and the infidelity is just $1-F$. The fidelity does, however, not completely
characterize the produced entangled state. Depending on the exact method used
to generate the entangled state, one can in some situations argue that the
error will predominantly be, e.g., only a phase error
\cite{Duan03,Simon03,Barrett05}, whereas in other situations it will be a
combination of phase and bit flip errors (see \cite{Childress05} and
references therein). Below we shall therefore both consider the situation
where we only have a dephasing error as well the situation, where we have a
more complicated depolarizing error (exact definition given later). As we
shall see, the knowledge that the error is of a particular type (e.g. only
dephasing error) provides a significant advantage for purification.

We will also assume a separation of error probabilities: any internal, unitary
operation within the register fails with extremely low probability, $p_{L}$,
while all operations connecting the communication qubit to the outside world
(initialization, measurement, and entanglement generation) fail with error
probabilities that can be several orders of magnitude higher.%
\begin{equation}
p_{L}\ll p_{I},p_{M},1-F.
\end{equation}
In terms of these quantities the error probability in the non-local CNOT gate
in Fig.~\ref{fig:CNOT} is
\begin{equation}
p_{CNOT}\sim(1-F)+2p_{L}+2p_{M}. , \label{eq:pCNOT}%
\end{equation}
because we use one entangled state, two local operations, and two
measurements. In the next two sections, we will show how to use robust
operations to dramatically improve the fidelity for these non-local coupling gates.

\emph{Robust measurement} can be implemented by repeated QND measurement,
i.e., a majority vote among the measurement outcomes
(Fig.~\ref{fig:MeasureCircuit}), following a sequence of CNOT operations
between the auxiliary/storage qubit and the communication qubit. This also
allows \emph{robust initialization} by measurement. High-fidelity,
\emph{robust entanglement generation} is achieved via entanglement pumping
\cite{Briegel98,Dur99,Dur03} (Figs.~\ref{fig:ENPCartoon} and
\ref{fig:ENPCircuit}), in which lower fidelity entanglement between the
communication qubits is used to purify entanglement between the auxiliary
qubits, which can then be used for non-local CNOT operations. To make the most
efficient use of physical qubits, we introduce a new entanglement pumping
scheme. In our \emph{bit-phase two-level entanglement pumping} scheme, we
first use unpurified Bell pairs to repeatedly pump (purify) against bit-errors
(Fig.~\ref{fig:ENPCircuit}a), and then use the bit-purified Bell pairs to
repeatedly pump against phase-errors (Fig.~\ref{fig:ENPCircuit}b).

Entanglement pumping, like entanglement generation, is probabilistic; however,
failures are detected. In computation, where each logical gate should be
completed within the allocated time (clock cycle), failed entanglement pumping
can lead to gate failure. To demonstrate the feasibility of our approach for
quantum computation, we will analyze the time required for robust
initialization, measurement and entanglement generation, and show that the
failure probability for these procedures can be made sufficiently small with
reasonable time overhead.

\section{Robust Measurement \&
Initialization\label{Robust Measurement & Initialization}}

\begin{figure}[tb]
\begin{center}
\includegraphics[
width=8.5 cm
]{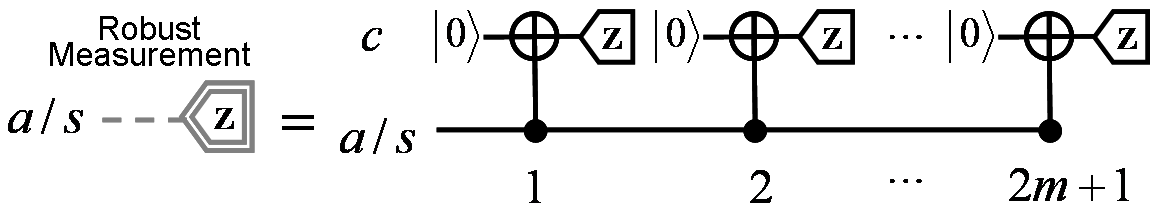}
\end{center}
\caption[fig:MeasureCircuit]{The robust measurement scheme based on repeated
quantum non-demolition (QND) measurements and majority vote. Each
QND\ measurement consists of initializing, coupling, and measuring the
communication qubit. The QND measurements are repeated $2m+1$ times using the
same communication qubit.}%
\label{fig:MeasureCircuit}%
\end{figure}

In this section, we will analyze the robust measurement scheme based on
repeated QND\ measurement, discuss the recent experimental demonstration of
robust measurement in the ion-trap system, and present two approaches to
robust initialization.

The measurement circuit shown in Fig.~\ref{fig:MeasureCircuit} yields the
correct result based on a majority vote from $2m+1$ consecutive readouts
(bit-verification). Since the evolution of the system (CNOT gate) commutes
with the measured observable ($Z$ operator) of the auxiliary/storage qubit, it
is a quantum non-demolition (QND) measurement \cite{Braginsky80}, which can be
repeated many times. The error probability for such a majority vote
measurement scheme is:%
\begin{equation}
\varepsilon_{M}\approx\sum_{j=m+1}^{2m+1}\left(
\begin{array}
[c]{c}%
2m+1\\
j
\end{array}
\right)  \left(  p_{I}+p_{M}\right)  ^{j}+\frac{2m+1}{2}p_{L} \label{eq:em}%
\end{equation}
where the last term conservatively estimates the probability for bit-flip
error of the auxiliary/storage qubit during the repeated QND\ measurement. For
simplicity, we will use Eq.~(\ref{eq:em}) for our conservative estimate of
error probability for repeated QND\ measurement. Suppose $p_{I}=p_{M}=5\%$, we
can achieve $\varepsilon_{M}\approx8\times10^{-4}$ by choosing $m^{\ast}=6$
for $p_{L}=10^{-4}$, or even $\varepsilon_{M}\approx12\times10^{-6}$ by
choosing $m^{\ast}=10$ for $p_{L}=10^{-6}$. For convenience of discussion, we
shall add $\varepsilon_{M}$ to the set of imperfection parameters: $\left(
1-F,p_{I},p_{M},p_{L},\varepsilon_{M}\right)  $. The time for robust
measurement is%
\begin{equation}
\tilde{t}_{M}=\left(  2m+1\right)  \left(  t_{I}+t_{L}+t_{M}\right)  .
\label{eq:tm}%
\end{equation}
where $t_{I}$, $t_{L}$, and $t_{M}$ are times for initialization, local
unitary gate, and measurement, respectively.

Measurements with very high fidelity ($\varepsilon_{M}$ as low as
$6\times10^{-4}$) have recently been demonstrated in the ion-trap system
\cite{Hume07}, using similar ideas as above. There are several possibilities
to further improve the performance of the repeated QND measurement. (1) We may
use maximum likelihood estimate (MLE) to replace the majority vote for
repeated measurements with multi-value outcomes (e.g., fluorescent intensity)
\cite{Hume07}. (2) We may keep updating the error probability using MLE after
each measurement. Once the estimated error probability is below some fixed
error rate, we stop the repetition of the QND measurement to avoid errors from
redundant operations \cite{Hume07}. (3) We may use the implementation of CNOT
gate that has small/vanishing bit-flip errors to the control qubit, which will
reduce/eliminate the last term in Eq.~(\ref{eq:em}).

The robust measurement scheme also allows to achieve robust initialization by
measurement, i.e., by measuring the state of a qubit with the robust
measurement scheme, we initialize the qubit into the result of the measurement
outcome with an effective initialization error%
\begin{equation}
\varepsilon_{I}\approx\varepsilon_{M}. \label{eq:ei1}%
\end{equation}
Besides the above measurement-based scheme, we may achieve robust
initialization using verification-based scheme \footnote[50]{The
verification-based scheme requires $k\geq0$ verifications for a good
initialization and suppresses the initialization error to $\varepsilon
_{I}\approx\left(  p_{I}+p_{M}\right)  ^{k+1}+\frac{2k+1}{2}p_{L}$, a smaller
error probability than that of the measurement-based scheme. However, the
verification-based scheme may fail at some intermediate step and require
reiteration of the verification from scratch, which is very similar to the
one-level entanglement pumping, described by the Markov chain model in Sec.
\ref{Markov Chain Model}.}. For clarity, we shall assume the measurement-based
initialization [Eq.~(\ref{eq:ei1})] for the rest of the paper.

\section{Robust Non-Local Two-Qubit
Gate\label{Robust Non-Local Two-Qubit Gate}}

\begin{figure}[tb]
\begin{center}
\includegraphics[
width=8 cm
]{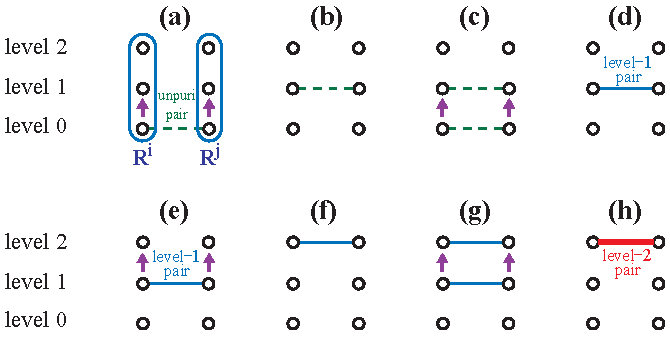}
\end{center}
\caption[fig:ENPCartoon]{(Color online) Two-level entanglement pumping between
registers $R^{i}$ and $R^{j}$ [circled by rounded rectangles in (a)]. (a,b)
Generate and store one unpurified Bell pair. (c) Generate another unpurified
Bell pair to pump (purify) the previously stored pair. (d) If the purification
is successful, we obtain a purified Bell pair (level-1 pair) with higher
fidelity; otherwise we discard the stored Bell pair and start the entire
pumping process from the beginning. (e-f) The second level of entanglement
pumping uses previously purified pairs to purify a stored Bell pair, to obtain
a Bell pair with higher fidelity (level-2 pair).}%
\label{fig:ENPCartoon}%
\end{figure}

With high-fidelity local unitary gate and repeated QND\ measurement, the error
probability for non-local coupling gates (e.g., Fig.~\ref{fig:CNOT} and
\ref{fig:Controlled-U}) is%
\begin{equation}
p_{CNOT}\sim(1-F)+2p_{L}+2\varepsilon_{M},
\end{equation}
which is dominated by the infidelity of the Bell pair $1-F$, since we assume
$p_{L}\sim\varepsilon_{M}\ll1-F$. In this section, we will show how to create
high-fidelity Bell pairs between two registers with a reasonable time
overhead. We will first briefly review entanglement pumping
\cite{Briegel98,Dur99,Childress05}. Then we will quantitatively analyze the
fidelity of the purified Bell pairs for our efficient two-level pumping
scheme, and introduce the Markov chain model to calculate the failure
probability for entanglement pumping within a given number of attempts. Next
we will quantify the performance of the high-fidelity Bell pair generation in
terms of the total error probability (or average infidelity) and the time
overhead, and discuss the trade-off between these two criteria. Finally, we
will mention a non-post-selective pumping scheme which may further reduce the
time overhead.

\subsection{Entanglement pumping}

We now consider entanglement pumping \cite{Briegel98,Dur99,Childress05} with
high-fidelity local unitary gate and robust measurement. During the
entanglement pumping process, we first store one unpurified Bell pair
[Fig.~\ref{fig:ENPCartoon} (a,b)], and then generate another unpurified Bell
pair to purify the previously stored pair [Fig.~\ref{fig:ENPCartoon} (c,d)].
If the purification is successful, we will obtain a purified Bell pair with
higher fidelity, which can be further purified by repeating the process in
Fig.~\ref{fig:ENPCartoon} (c,d) with new unpurified Bell pairs; otherwise we
discard the stored Bell pair, and start the entire pumping process from the
beginning. Sometimes, we may want to introduce a second level of entanglement
pumping; that is to use previously purified pairs to purify a stored Bell pair
[Fig.~\ref{fig:ENPCartoon} (e-h)].

\subsection{Fidelity of entanglement pumping}

\begin{figure}[tb]
\begin{center}
\includegraphics[
width=8.5 cm
]{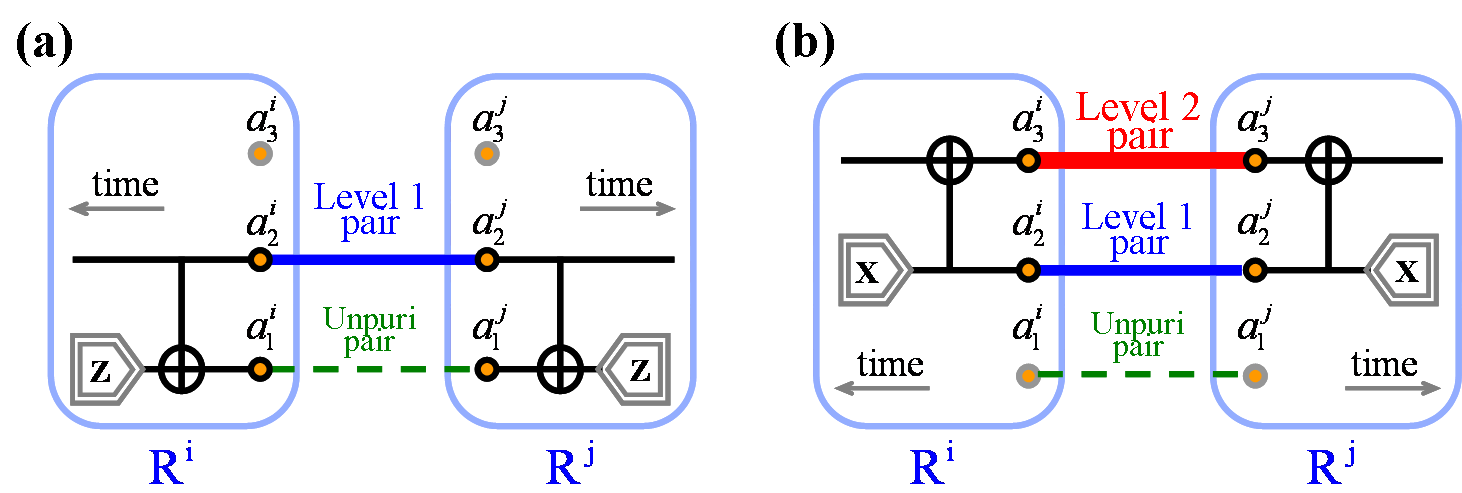}
\end{center}
\caption[fig:ENPCircuit]{(Color online) Bit-phase two-level entanglement
pumping scheme to create high fidelity entangled pairs between two registers
$R^{i}$ and $R^{j}$. (a) Circuit for the first level pumping to purify
bit-errors, corresponding to Fig.~\ref{fig:ENPCartoon}c. (b) Circuit for the
second level pumping to purify phase-errors, corresponding to
Fig.~\ref{fig:ENPCartoon}g. The arrows indicate the time direction for each
register. Robust measurements are used here. If the two outcomes are the same,
it is a successful attempt of pumping; otherwise we generate new pairs and
restart the pumping operation from the beginning. }%
\label{fig:ENPCircuit}%
\end{figure}

\begin{figure}[tb]
\begin{center}
\includegraphics[
width=8.7 cm
]{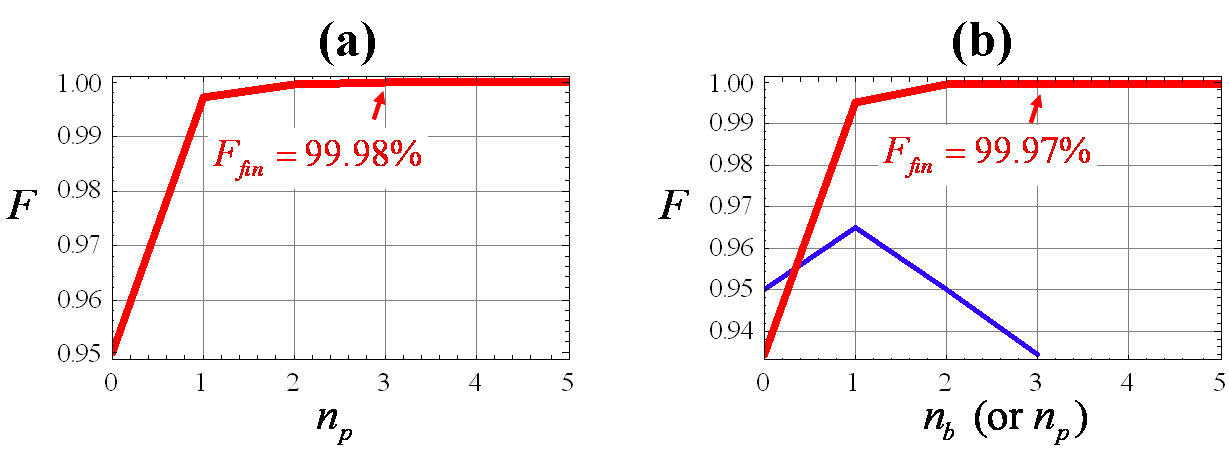}
\end{center}
\caption[fig:ENPCurve]{(Color online) Entanglement fidelity $F$ as a function
of the number of successful pumping steps. $n_{b}$ and $n_{p}$ are
the number of pumping steps used to purify bit-errors and phase-errors,
respectively. We assume fixed measurement and local two-qubit
gate error rates $\varepsilon_{M}=p_{L}=10^{-4}$. (a) For bit-flip error
$\rho_{\text{\textrm{dephasing}}}=diag\left[  F,1-F,0,0\right]  $ in the Bell
basis $\left\{  \left\vert \Phi^{+}\right\rangle ,\left\vert \Phi
^{-}\right\rangle ,\left\vert \Psi^{+}\right\rangle ,\left\vert \Psi
^{-}\right\rangle \right\}  $, with $F=0.95$, one-level entanglement pumping
is sufficient. High fidelity of $F_{fin}=99.98\%$ can be achieved by $n_{p}%
=3$. (b) For depolarizing error $\rho_{\mathrm{depolarizing}}=diag\left[
F,\frac{1-F}{3},\frac{1-F}{3},\frac{1-F}{3}\right]  $, with $F=0.95$,
two-level entanglement pumping is needed (see text for more details). The
first level pumping only purifies the bit-error, but accumulates the
phase-error at the same time, and therefore the (thin blue) fidelity curve for
the first level pumping drops for $n_{b}>1$. The second level (thick red
curve) uses the purified level-1 pair ($n_{b}=1$) to pump another stored pair.
High fidelity of $F_{fin}=99.97\%$ can be achieved by $n_{p}=3$. Note that we
consider the full density matrix for all numerical calculation of entanglement
fidelities \cite{Dur98}, with the error models given in Eqs.
(\ref{eq:ErrorModel1},\ref{eq:ErrorModel2},\ref{eq:ErrorModel3}).}%
\label{fig:ENPCurve}%
\end{figure}

We now analyze the performance of entanglement pumping for different errors of
the unpurified Bell pairs. If the unpurified Bell pair is dominated by one
type of error (e.g., dephasing error with density matrix $\rho
_{\text{\textrm{dephasing}}}=diag\left[  F,1-F,0,0\right]  $ in the Bell basis
$\left\{  \left\vert \Phi^{+}\right\rangle ,\left\vert \Phi^{-}\right\rangle
,\left\vert \Psi^{+}\right\rangle ,\left\vert \Psi^{-}\right\rangle \right\}
$, defined as $\left\vert \Phi^{\pm}\right\rangle =\left(  \left\vert
00\right\rangle \pm\left\vert 11\right\rangle \right)  /\sqrt{2}$, $\left\vert
\Psi^{\pm}\right\rangle =\left(  \left\vert 01\right\rangle \pm\left\vert
10\right\rangle \right)  /\sqrt{2}$), we can skip the first level pumping. The
unpurified pair then immediately becomes a level-1 pair and is purified with
the circuit in Fig.~\ref{fig:ENPCircuit}b. In Fig.~\ref{fig:ENPCurve}a we plot
the fidelity curve (purified fidelity v.s. number of successful pumping steps)
for the one-level pumping process (i.e. $n_{b}=0$), where a very high fidelity
pair can be created after $n_{p}=3$ successful pumping steps. Note that we
consider the full density matrix for all numerical calculation of entanglement
fidelities \cite{Dur98}, with the error models given in Eqs.
(\ref{eq:ErrorModel1},\ref{eq:ErrorModel2},\ref{eq:ErrorModel3}).

If the unpurified Bell pair contains errors from both bit-flip and dephasing
processes (e.g., depolarizing error with density matrix $\rho
_{\mathrm{depolarizing}}=diag\left[  F,\frac{1-F}{3},\frac{1-F}{3},\frac
{1-F}{3}\right]  $ in the Bell basis), two-level entanglement pumping is
needed. We introduce the following \emph{bit-phase two-level pumping
scheme}\label{sec:two-level pumping scheme} ---the first level has $n_{b}$
steps of bit-error pumping using raw Bell pairs (Fig.~\ref{fig:ENPCircuit}a)
to produce a bit-error-purified entangled pair, and the second level uses
these bit-error-purified pairs for $n_{p}$ steps of phase-error pumping
(Fig.~\ref{fig:ENPCircuit}b). In Fig.~\ref{fig:ENPCurve}b we plot the fidelity
curves for the first level (thin blue curve) and the second level (thick red
curve) of entanglement pumping. One-level pumping is insufficient to achieve
high fidelity, but two-level pumping can achieve very high fidelity. With the
parameters specified for Fig.~\ref{fig:ENPCurve}b,\ the maximum fidelity is
achieved via the optimal choice of control parameters $\left(  n_{b}^{\ast
},n_{p}^{\ast}\right)  =\left(  1,3\right)  $ for successful pumping steps of
the first and second levels, respectively.

For successful purification, the infidelity of the purified pair,
$\varepsilon_{E,\text{\textrm{infid}}}^{\left(  n_{b},n_{p}\right)  }$,
depends on both the control parameters $\left(  n_{b},n_{p}\right)  $ and the
imperfection parameters $\left(  F,p_{L},\varepsilon_{M}\right)  $. For
depolarizing error, we find%
\begin{align}
&  \varepsilon_{E,\text{\textrm{infid}}}^{\left(  n_{b}\geq1,n_{p}%
\geq1\right)  }\approx\frac{3+2n_{p}}{4}p_{L}+\frac{4+2\left(  n_{b}%
+n_{p}\right)  }{3}\left(  1-F\right)  \varepsilon_{M}\\
&  +\left(  n_{p}+1\right)  \left(  \frac{2\left(  1-F\right)  }{3}\right)
^{n_{b}+1}+\left(  \frac{\left(  n_{b}+1\right)  \left(  1-F\right)  }%
{3}\right)  ^{n_{p}+1}\nonumber
\end{align}
to leading order in $p_{L}$ and $\varepsilon_{M}$, for $n_{b},n_{p}\geq1$. The
dependence on the initial infidelity $1-F$ is exponentially suppressed at a
cost of a linear increase of error from local operations $p_{L}$ and robust
measurement $\varepsilon_{M}$. Measurement-related errors are suppressed by
the prefactor $1-F$, since measurement error does not cause infidelity unless
combined with other errors. In the limit of ideal operations ($p_{L}%
,\varepsilon_{M}\rightarrow0$), the infidelity $\varepsilon
_{E,\text{\textrm{infid}}}^{\left(  n_{b},n_{p}\right)  }$ can be arbitrarily
close to zero, which is rigorously proved in Appendix
\ref{Bit-phase two-level pumping scheme}. On the other hand, if we use the
standard entanglement pumping scheme \cite{Briegel98,Dur99} (that alternates
purification of bit and phase errors within each pumping level), the reduced
infidelity from two-level pumping is always larger than $\left(  1-F\right)
^{2}/9$. Therefore, for very small $p_{L}$ and $\varepsilon_{M}$, the new
pumping scheme is crucial to minimize the number of qubits per register.

\begin{figure}[tb]
\begin{center}
\includegraphics[
width=8.5 cm
]{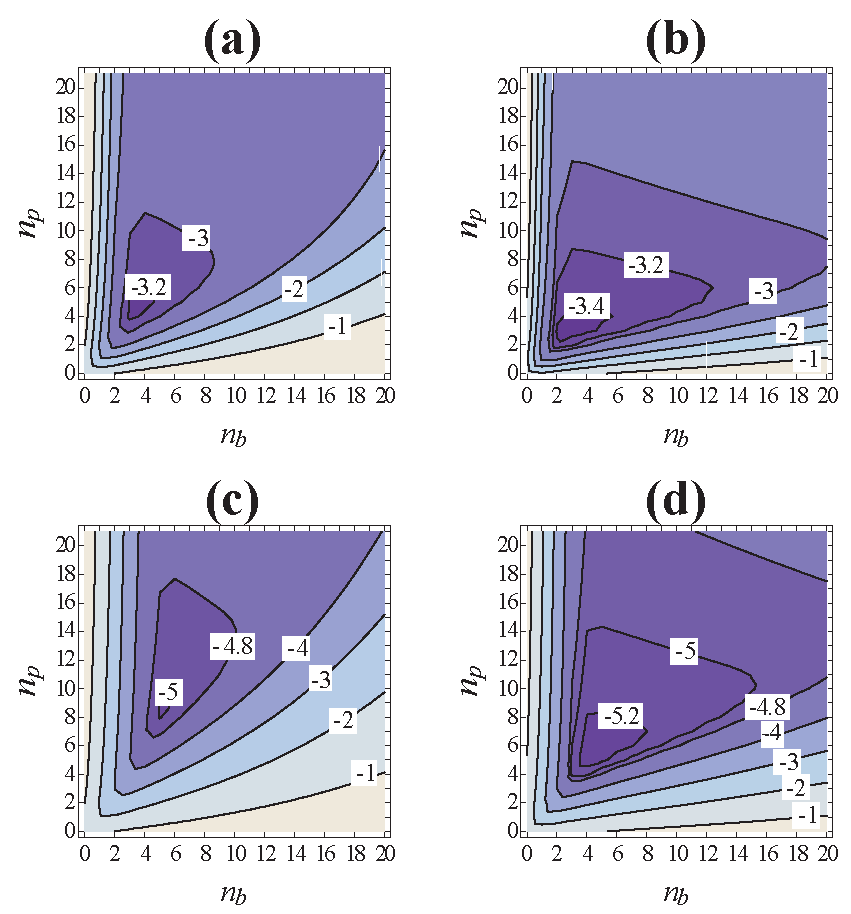}
\end{center}
\caption[fig:ENPInfidContour]{(Color online) The contours of the infidelity
$\varepsilon_{E,\text{\textrm{infid}}}^{\left(  n_{b},n_{p}\right)  }$ as a
function of $n_{b}$ and $n_{p}$ for depolarizing error. We use $\log
_{10}\varepsilon_{E,\text{\textrm{infid}}}^{\left(  n_{b},n_{p}\right)  }$ to
label the contours. The other parameters are $F=0.90$ (left) and $F=0.95$
(right); $\tilde{\varepsilon}_{M}=p_{L}=10^{-4}$ (up) and $\tilde{\varepsilon
}_{M}=p_{L}=10^{-6}$ (down). With optimal choice of $\left(  n_{b}^{\ast
},n_{p}^{\ast}\right)  $, the minimal infidelity is comparable to the
corresponding value of $p_{L}$. }%
\label{fig:ENPInfidContour}%
\end{figure}

In Fig.~\ref{fig:ENPInfidContour}, we show the contours of the infidelity
$\varepsilon_{E,\text{\textrm{infid}}}^{\left(  n_{b},n_{p}\right)  }$ as a
function of $n_{b}$ and $n_{p}$, where the contours are labeled by values of
$\log_{10}\varepsilon_{E,\text{\textrm{infid}}}^{\left(  n_{b},n_{p}\right)
}$. The parameters for the contour plots are $F=0.90$ (left) and $F=0.95$
(right); $\varepsilon_{M}=p_{L}=10^{-4}$ (up) and $\varepsilon_{M}%
=p_{L}=10^{-6}$ (down). For optimal choice of $\left(  n_{b}^{\ast}%
,n_{p}^{\ast}\right)  $, the minimal infidelity is limited by $\varepsilon
_{M}$ and $p_{L}$.

For dephasing error, one level pumping is sufficient (i.e. no bit-error
purification, $n_{b}=0$). The infidelity is approximately
\begin{equation}
\varepsilon_{E}^{\left(  0,n_{p}\geq1\right)  }\approx\left(  1-F\right)
^{n_{p}+1}+\frac{2+n_{p}}{4}p_{L}+2\left(  1-F\right)  \varepsilon_{M}%
\end{equation}
by expanding to the leading order in $p_{L}$ and $\varepsilon_{M}$.

\section{Markov Chain Model \label{Markov Chain Model}}

The overall success probability can be defined as the joint probability that
all successive steps succeed. We use the model of finite-state Markov chain
\cite{Meyn93} to directly calculate the \emph{failure probability} of $\left(
n_{b},n_{p}\right)  $-two-level entanglement pumping using $N_{\mathrm{tot}}$
raw Bell pairs, denoted as $\varepsilon_{E,fail}^{\left(  n_{b},n_{p}\right)
}\left(  N_{\mathrm{tot}}\right)  $.

\subsection{Markov chain model for entanglement pumping}

\begin{figure}[tb]
\begin{center}
\includegraphics[
height=4.5 cm,
]{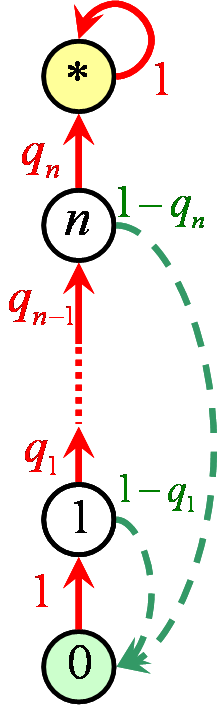}
\end{center}
\caption[fig:MarkovChain1]{(Color online) Markov chain model for one-level
entanglement pumping. We use "$0$" to denote the \emph{initial state} with no
Bell pairs, "$1$" for the state with one stored unpurified pair, $\left(
j+1\right)  $ for the state with one purified pair surviving $j$ steps of
pumping, and "$\ast$" for the \emph{final state} with the purified pair
surviving $n$ steps of pumping. The (success) transition probability from
state $j$ to state $j+1$ is $q_{j}$, while the (failure) transition
probability from state $j$ to state $0$ is $1-q_{j}$, for $j=0,1,\cdots,n$.
Here $q_{0}\equiv1$ and $q_{j\geq1}$ can be calculated according to the
density matrix of the purified Bell pair surviving $j$ steps of pumping
\cite{Dur99}. The final state is self-trapped, and goes back to itself with
unit probability. Each transition attempt consumes one unpurified Bell pair.}%
\label{fig:MarkovChain1}%
\end{figure}

We first use the Markov chain model to study the $n$-step one-level
entanglement pumping. As shown in Fig.~\ref{fig:MarkovChain1}, we use "$0$" to
denote the \emph{initial state} with no Bell pairs, "$1$" for the state with
one stored unpurified pair, $\left(  j+1\right)  $ for the state with one
purified pair surviving $j$ steps of pumping, and "$\ast$" for the \emph{final
state} with the purified pair surviving $n$ steps of pumping. Altogether there
are $n+2$ states. The (success) transition probability from state $j$ to state
$j+1$ is $q_{j}$, while the (failure) transition probability from state $j$ to
state $0$ is $1-q_{j}$, for $j=0,1,\cdots,n$. Here $q_{0}\equiv1$
[corresponding to deterministic state transfer as shown in
Fig.~\ref{fig:ENPCartoon} (a,b)] and $q_{j\geq1}$ can be calculated with the
density matrix of the purified Bell pair surviving $j-1$ steps of pumping
\cite{Dur99}. The final state is self-trapped, and goes back to itself with
unit probability, representing that once we have reached the desired final
fidelity we no longer make any purification attempts and the system remains in
this state with unit probability. Each transition attempt consumes one
unpurified Bell pair. We would like to know the probability of reaching the
final state "$\ast$" after $N_{\mathrm{tot}}$ attempts. More generally, we
might also want to know the probability distribution over all $n+2$ states.

We use a (column) vector $\vec{P}$ with $n+2$ elements to characterize the
probability distribution among all $n+2$ states. From the $t$-th attempt to
the $\left(  t+1\right)  $-th attempt, the probability vector evolves from
$\vec{P}_{t}$ to $\vec{P}_{t+1}$ according to the rule%
\begin{equation}
\vec{P}_{t+1}=\mathbf{M}~\vec{P}_{t}, \label{eq:ProbVecEvolve}%
\end{equation}
with the transition matrix%
\begin{equation}
\mathbf{M}=\left(
\begin{array}
[c]{cccccc}%
0 & 1-q_{1} & 1-q_{2} & \cdots & 1-q_{n} & 0\\
1 & 0 &  &  &  & \\
& q_{1} & 0 &  &  & \\
&  & q_{2} & 0 &  & \\
&  &  & \cdots & 0 & \\
&  &  &  & q_{n} & 1
\end{array}
\right)  . \label{eq:TranMatrix1}%
\end{equation}
Since the initial probability vector is $\vec{P}_{0}=\left(  1,0,\cdots
,0\right)  ^{T}$, we can calculate the probability vector after
$N_{\mathrm{tot}}$ attempts%
\begin{equation}
\vec{P}_{N_{\mathrm{tot}}}=\mathbf{M}^{N_{\mathrm{tot}}}\vec{P}_{0}.
\label{eq:ProbVecNtot}%
\end{equation}
The probability vector $\vec{P}_{N_{\mathrm{tot}}}$ describes the entire
probability distribution over all states of the Markov chain. The last element
of $\vec{P}_{N_{\mathrm{tot}}}$ is the success probability of reaching the
final state "$\ast$" after $N_{\mathrm{tot}}$ attempts; the failure
probability after $N_{\mathrm{tot}}$ attempts is thus%
\begin{equation}
\varepsilon_{E,fail}^{\left(  n_{b},n_{p}\right)  }\left(  N_{\mathrm{tot}%
}\right)  =1-\left(  P_{N_{\mathrm{tot}}}\right)  _{n+2}\text{.}%
\end{equation}

\begin{figure}[tb]
\begin{center}
\includegraphics[
width=8.5 cm
]{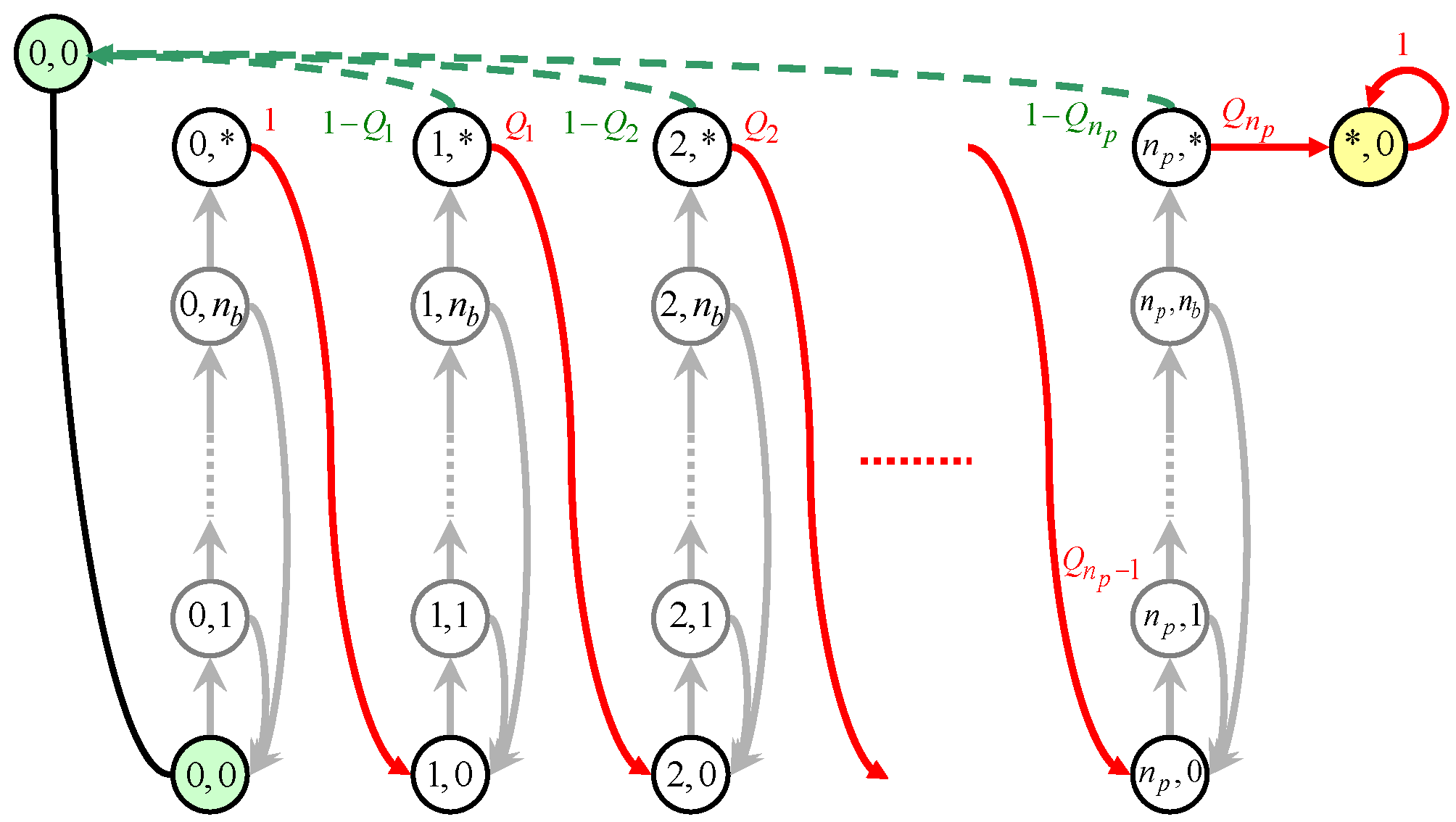}
\end{center}
\caption[fig:MarkovChain2]{(Color online) Markov chain model for two-level
entanglement pumping. The required pumping steps are $n_{b}$ and $n_{p}$ for
the two levels, respectively. We use "$0,0$" to denote the initial state with
no Bell pairs, "$0,j+1$" for the state with one purified pair surviving $j$
steps of pumping at the first level, "$k+1,j+1$" for the state with one
purified pair surviving $k$ steps of pumping at the second level and one
purified pair surviving $j$ steps of pumping at the first level, "$k+1,\ast$"
for the state with one purified pair surviving $k$ steps of pumping at the
second level and one purified pair surviving $n_{b}$ steps of pumping at the
first level, and "$\ast,0$" for the final state with one purified pair
surviving $n_{p}$ steps of pumping at the second level. For the first level
pumping, the (success) transition probability from state "$k,j$" to state
"$k,j+1$" is $q_{j}$, while the (failure) transition probability from state
"$k,j$" to state "$k,0$" is $1-q_{j}$, for $j=0,1,\cdots,n_{b}$. For the
second level pumping, the (success) transition probability from state
"$k,\ast$" to state "$k+1,0$" is $Q_{k}$, while the (failure) transition
probability from state "$k,\ast$" to state "$0,0$" is $1-Q_{k}$, for
$k=0,1,\cdots,n_{p}$. The final state is self-trapped, and goes back to itself
with unit probability.}%
\label{fig:MarkovChain2}%
\end{figure}

For two-level entanglement pumping, the state transition diagram is shown in
Fig.~\ref{fig:MarkovChain2}. $n_{b}$ and $n_{p}$ are the number of pumping
steps used to purify bit-errors and phase-errors, respectively. As detailed in
Appendix \ref{Markov Chain Model for Two-Level Pumping}, we may use a (column)
vector $\vec{P}$ with $\left(  n_{b}+1\right)  \left(  n_{p}+1\right)  +1$
elements to characterize the probability distribution among all $\left(
n_{b}+1\right)  \left(  n_{p}+1\right)  +1$ states. From the $t$-th attempt to
the $\left(  t+1\right)  $-th attempt, the probability vector evolves from
$\vec{P}\left(  t\right)  $ to $\vec{P}\left(  t+1\right)  $ according to the
same rule as above [Eq.~(\ref{eq:ProbVecEvolve})], but with the transition
matrix $\mathbf{M}$ given in Eq.~(\ref{eq:TranMatrix2}).

Similar to one-level pumping, we can calculate the probability vector after
$N_{\mathrm{tot}}$ attempts using Eq.~(\ref{eq:ProbVecNtot}). The probability
vector $\vec{P}_{N_{\mathrm{tot}}}$ describes the entire probability
distribution over all states of the Markov chain. The last element of $\vec
{P}_{N_{\mathrm{tot}}}$ is the success probability of reaching the final state
"$\ast,0$" after $N_{\mathrm{tot}}$ attempts; the failure probability after
$N_{\mathrm{tot}}$ attempts is then
\begin{equation}
\varepsilon_{E,\text{\textrm{fail}}}^{\left(  n_{b},n_{p}\right)  }\left(
N_{\mathrm{tot}}\right)  =1-P\left(  N_{\mathrm{tot}}\right)  _{\left(
n_{b}+1\right)  \left(  n_{p}+1\right)  +1}\text{.}%
\end{equation}

\begin{figure}[tb]
\begin{center}
\includegraphics[
width=7 cm
]{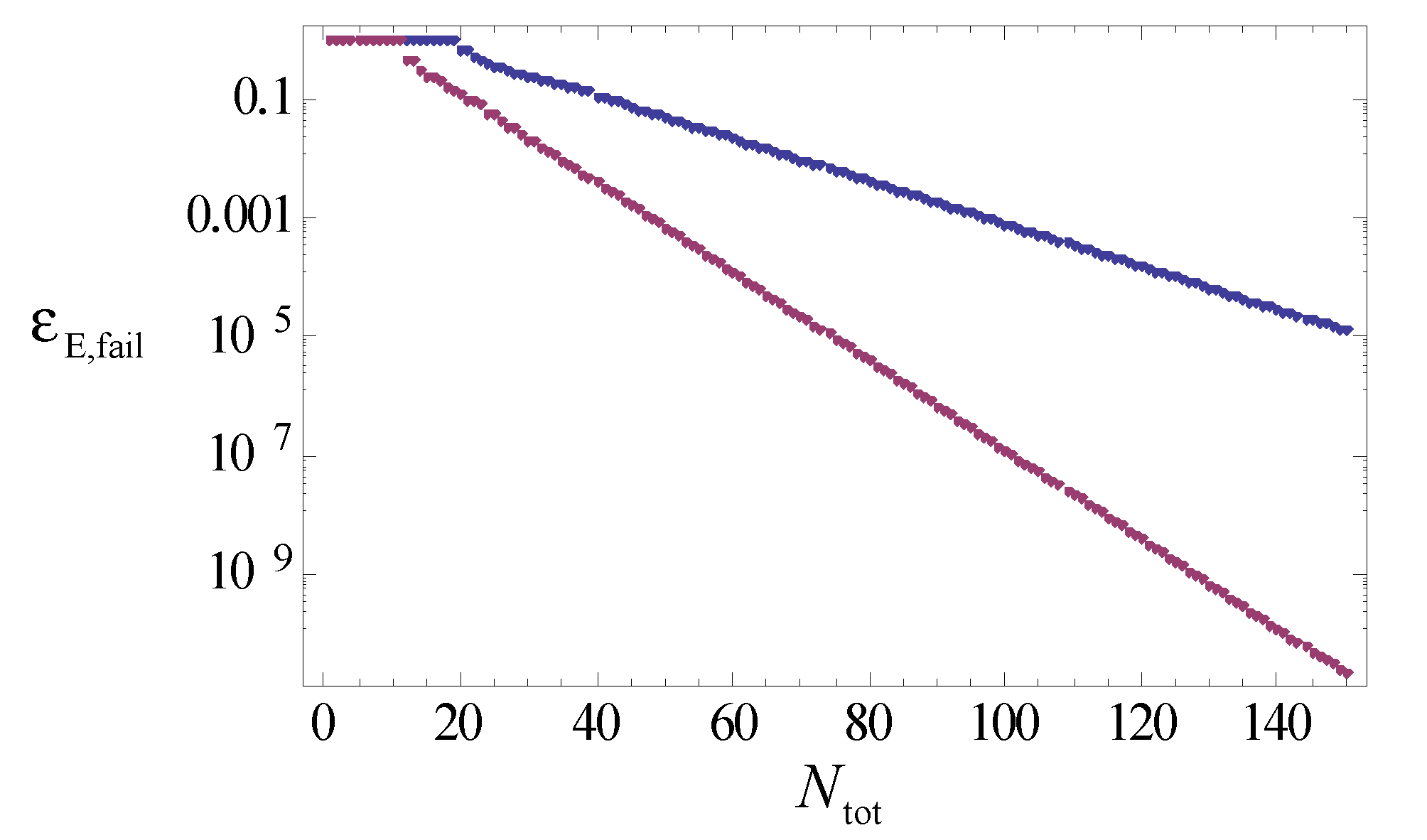}
\end{center}
\caption[fig:ENPFailureProb]{(Color online) Failure probability $\varepsilon
_{E,\text{\textrm{fail}}}$ as a function of $N_{\mathrm{tot}}$. We assume a
depolarizing error with $F=0.95$, $p_{L}=\tilde{\varepsilon}_{M}=10^{-4}$. We
choose $\left(  n_{b},n_{p}\right)  =\left(  2,3\right)  $ for the lower
curve, and $\left(  3,4\right)  $ for the upper curve. For large
$N_{\mathrm{tot}}$, the failure probability $\varepsilon
_{E,\text{\textrm{fail}}}$ decreases exponentially with $N_{\mathrm{tot}}$.}%
\label{fig:ENPFailureProb}%
\end{figure}

In Fig.~\ref{fig:ENPFailureProb}, we plot the failure probability
$\varepsilon_{E,\text{\textrm{fail}}}^{\left(  n_{b},n_{p}\right)  }\left(
N_{\mathrm{tot}}\right)  $ v.s. $N_{\mathrm{tot}}$, for control parameters
$\left(  n_{b},n_{p}\right)  =\left(  2,3\right)  $ and $\left(  3,4\right)
$. For $N_{\mathrm{tot}}$ sufficiently large, the failure probability
decreases exponentially to zero. For any given parameters, we can efficiently
suppress the failure probability with some reasonably large $N_{\mathrm{tot}}$.

\subsection{Total error probability \& average infidelity}

We now introduce the \emph{total error probability} (TEP) approximated by the
sum of the failure probability and the infidelity of the purified Bell pair
\begin{equation}
\varepsilon_{E}^{\left(  n_{b},n_{p}\right)  }\left(  N_{\mathrm{tot}}\right)
\approx\varepsilon_{E,\text{\textrm{fail}}}^{\left(  n_{b},n_{p}\right)
}\left(  N_{\mathrm{tot}}\right)  +\varepsilon_{E,\text{\textrm{infid}}%
}^{\left(  n_{b},n_{p}\right)  }. \label{eq:DefTEP}%
\end{equation}
This is a very conservative estimate, since sometimes we do create some
partially purified Bell pair though not the targeted purified Bell pair. And
here we just say that the state has fidelity zero in these cases.

To consider the possibility of using a partially purified Bell pair for
output, we may introduce another useful quantity ---the \emph{average
infidelity} (AIF) ---for the output Bell pair from the robust entanglement
generation, where we take into account these partially purified pairs. The
average infidelity of the output pair is the weighted average of the
infidelity of the Markov chain%
\begin{align}
&  \delta_{E}^{\left(  n_{b},n_{p}\right)  }\left(  N_{\mathrm{tot}}\right)
\nonumber\\
&  \equiv1-\left\langle F_{N_{\mathrm{tot}}}^{\left(  n_{b},n_{p}\right)
}\right\rangle \\
&  =\sum_{n_{b}^{\prime}=0}^{n_{b}}\sum_{n_{p}^{\prime}=0}^{n_{p}}%
\varepsilon_{E,\text{\textrm{infid}}}^{\left(  n_{b}^{\prime},n_{p}^{\prime
}\right)  }P\left(  N_{\text{\textrm{tot}}}\right)  _{\left(  n_{b}+1\right)
n_{p}^{\prime}+n_{b}^{\prime}+2}+\frac{1}{2}P\left(  N_{\text{\textrm{tot}}%
}\right)  _{1}.\nonumber
\end{align}
Here the first term sums over all states of the Markov chain (except for the
initial one), each of which has at least one partially purified pair with
infidelity $\varepsilon_{E,\text{\textrm{infid}}}^{\left(  n_{b}^{\prime
},n_{p}^{\prime}\right)  }$ and probability $P\left(  N_{\text{\textrm{tot}}%
}\right)  _{\left(  n_{b}+1\right)  n_{p}^{\prime}+n_{b}^{\prime}+2}$; the
last term comes from the situation that none of the partially purified Bell
pairs remain after the last attempt of the entanglement purification and we
just use a classically correlated pair with infidelity $1/2$. Generally, the
average infidelity is smaller than the total error probability.

\begin{figure}[tb]
\begin{center}
\includegraphics[
width=8.7 cm
]{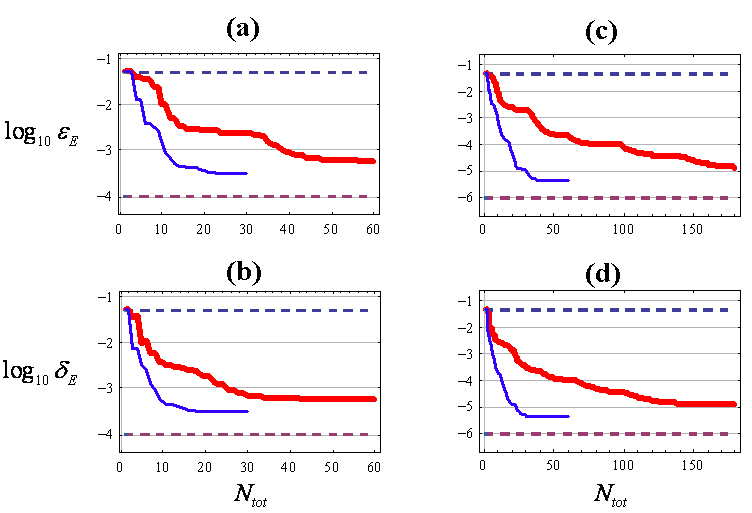}
\end{center}
\caption[fig:TEPandAIF]{(Color online) The optimized total error probability
$\varepsilon_{E}$ [Eq.~(\ref{eq:DefOptTEP})] (upper plots) and the optimized
average infidelity $\delta_{E}$ [Eq.~(\ref{eq:DefOptAIF})] (lower plots) as a
function $N_{\mathrm{tot}}$. The error probability for the local coupling
gates is $p_{L}=10^{-4}$ (left plots) and $p_{L}=10^{-6}$ (right plots).
One-level pumping is used for dephasing error (thin blue curves); two-level
pumping is used for depolarizing error (thick red curves). The other
parameters $F=0.95,p_{I}=p_{M}=5\%$ are the same for all plots. Both
$\varepsilon_{E}$ and $\delta_{E}$ saturate for large $N_{\mathrm{tot}}$. In
each plot, we also show the initial infidelity $1-F$ (upper blue dashed lines)
and the local error probability $p_{L}$ (lower violet dashed lines).}%
\label{fig:TEPandAIF}%
\end{figure}

We may also optimize the choice of the control parameters $\left(  n_{b}%
,n_{p}\right)  $%
\begin{equation}
\varepsilon_{E}\left(  N_{\mathrm{tot}}\right)  \equiv\min_{\left(
n_{b},n_{p}\right)  }\varepsilon_{E}^{\left(  n_{b},n_{p}\right)  }\left(
N_{\mathrm{tot}}\right)  \label{eq:DefOptTEP}%
\end{equation}
and%
\begin{equation}
\delta_{E}\left(  N_{\mathrm{tot}}\right)  \equiv\min_{\left(  n_{b}%
,n_{p}\right)  }\delta_{E}^{\left(  n_{b},n_{p}\right)  }\left(
N_{\mathrm{tot}}\right)  . \label{eq:DefOptAIF}%
\end{equation}
In Fig.~\ref{fig:TEPandAIF}, we plot both the optimized total error
probability $\varepsilon_{E}$ and the optimized average infidelity $\delta
_{E}$ as a function of $N_{\text{\textrm{tot}}}$. Both quantities
asymptotically approaches the same minimum value%
\begin{equation}
\lim_{N_{\mathrm{tot}}\rightarrow\infty}\varepsilon_{E}\left(  N_{\mathrm{tot}%
}\right)  =\lim_{N_{\mathrm{tot}}\rightarrow\infty}\delta_{E}\left(
N_{\mathrm{tot}}\right)  =\Delta_{\min}. \label{eq:AsymptoticMinimum}%
\end{equation}
Here the minimum value is simply the minimal infidelity of the entanglement
purification%
\begin{equation}
\Delta_{\min}\equiv\min_{\left(  n_{b},n_{p}\right)  }\varepsilon
_{E,\text{\textrm{infid}}}^{\left(  n_{b},n_{p}\right)  } \label{eq:DeltaMin}%
\end{equation}
which is achieved by the control parameters $\left(  n_{b},n_{p}\right)
\equiv\left(  n_{b}^{\ast},n_{p}^{\ast}\right)  $, for the imperfection
parameters $\left\{  p_{L},1-F,\varepsilon_{M}\right\}  $.

We remark that a faster and less resource intensive approach may be used if
the unpurified Bell pair is dominated by dephasing error. Then, one-level
pumping is sufficient (i.e. no bit-error purification, $n_{b}=0$). The
optimized total error probability and average infidelity (thin blue curves)
for this situation are plotted as a function of $N_{\mathrm{tot}}$ in
Fig.~\ref{fig:TEPandAIF}.

\subsection{Total time for robust entanglement generation}

The total time for robust entanglement generation $\tilde{t}_{E}$ is
proportional to the average number of raw Bell pairs generated $\left\langle
N_{\mathrm{tot}}\right\rangle $
\begin{equation}
\tilde{t}_{E}\approx\left\langle N_{\mathrm{tot}}\right\rangle \times\left(
t_{E}+t_{L}+\tilde{t}_{M}\right)  , \label{eq:te}%
\end{equation}
where $t_{E}$ is the average generation time of the unpurified Bell pair. Note
that the entanglement generation itself is a stochastic process. In principle,
we may also include the stochastic nature of the entanglement generation by
introducing a sub-level of Markov chain to characterize the stochastic
entanglement generation. Since each entanglement generation either succeeds or
fails, the sub-level Markov chain only involves two states, which can be
easily incorporated into the Markov chain models discussed above. After
incorporating the sub-level into the Markov chain, each transition corresponds
to one attempt of entanglement generation, instead of one attempt of
entanglement purification that consumes one unpurified Bell pair previously.

Nevertheless, the number of Bell pairs generated in a given period of time
(i.e. $N_{\mathrm{tot}}$) has a distribution. Since the relative deviation of
this distribution ($\sim N_{\mathrm{tot}}^{-1/2}$) is fairly small for large
$N_{\mathrm{tot}}$ ($>20$), this only has a minor influence. Thus we replace
$\left\langle N_{\mathrm{tot}}\right\rangle $ by $N_{\mathrm{tot}}$.

\begin{figure*}[tb]
\begin{center}
\includegraphics[
width= 13 cm
]{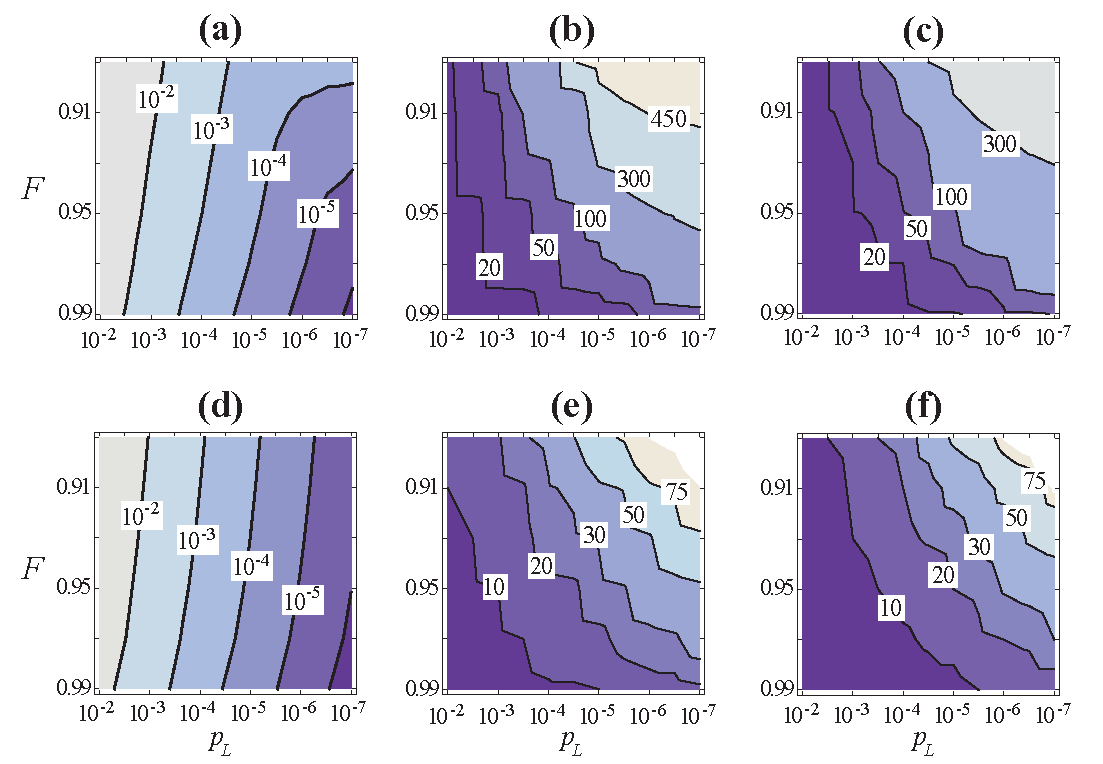}
\end{center}
\caption[fig:Contours4PS]{(Color online) Contours of the total error
probability $\varepsilon_{E}$ (or average infidelity $\delta_{E}$) after
purification (left), the total number of unpurified Bell pairs
$N_{\mathrm{tot}}$ associated with $\varepsilon_{E}$ [Eq.~(\ref{eq:TEP2Delta}%
)] (middle), and $N_{\mathrm{tot}}$ associated with $\delta_{E}$
[Eq.~(\ref{eq:AIF2Delta})] (right). The contours are drawn with respect to the
imperfection parameters $p_{L}$ (horizontal axis) and $F$ (vertical axis).
Two-level pumping (up) is used for depolarizing error, and one-level pumping
(down) for dephasing error. $p_{I}=p_{M}=5\%$ is assumed. }%
\label{fig:Contours4PS}%
\end{figure*}

\subsection{Trade-off between gate quality and time overhead}

We now consider the balance between the "quality" of the robustly generated
entangled pairs and the time overhead $N_{\mathrm{tot}}$ associated with the
robust generation process. We may use either the optimized total error
probability $\varepsilon_{E}\left(  N_{\mathrm{tot}}\right)  $ or the
optimized average infidelity $\delta_{E}\left(  N_{\mathrm{tot}}\right)  $ to
characterize the quality. Since both quantities approaches the same asymptotic
minimum $\Delta_{\min}$ according to Eq.~(\ref{eq:AsymptoticMinimum}), there
is only little improvement in the quality of the robust entanglement
generation once $\varepsilon_{E}\left(  N_{\mathrm{tot}}\right)  $ or
$\delta_{E}\left(  N_{\mathrm{tot}}\right)  $ is comparable to $\Delta_{\min}$
(say $2\Delta_{\min}$). Thus, we find the value for $N_{\mathrm{tot}}$ by
imposing the relations%
\begin{equation}
\varepsilon_{E}\left(  N_{\mathrm{tot}}\right)  =2\Delta_{\min}
\label{eq:TEP2Delta}%
\end{equation}
or%
\begin{equation}
\delta_{E}\left(  N_{\mathrm{tot}}\right)  =2\Delta_{\min}.
\label{eq:AIF2Delta}%
\end{equation}

First, we consider the total error probability $\varepsilon_{E}\left(
N_{\mathrm{tot}}\right)  $. The relation in Eq.~(\ref{eq:TEP2Delta}) can be
simplified, if we assume fixed control parameters $\left(  n_{b},n_{p}\right)
\equiv\left(  n_{b}^{\ast},n_{p}^{\ast}\right)  $ for the left hand side
(rather than minimizing over all possible choices of $\left(  n_{b}%
,n_{p}\right)  $). Combined with Eq.~(\ref{eq:DefTEP}), the failure
probability should be comparable to the minimal infidelity%
\begin{equation}
\varepsilon_{E,\text{\textrm{fail}}}^{\left(  n_{b}^{\ast},n_{p}^{\ast
}\right)  }\left(  N_{\mathrm{tot}}\right)  \approx\Delta_{\min}\text{.}
\label{eq:FailProb&Delta}%
\end{equation}
Since both the variable $\Delta_{\min}$ and the parameters $\left(
n_{b}^{\ast},n_{p}^{\ast}\right)  $ depend on $\left\{  p_{L},p_{I}%
,p_{M},1-F\right\}  $, the above relation implicitly determines
$N_{\mathrm{tot}}$ as a function a function of $\left\{  p_{L},p_{I}%
,p_{M},1-F\right\}  $.

In Fig.~\ref{fig:Contours4PS}, we plot the contours of $\varepsilon_{E}$
[Eq.~(\ref{eq:TEP2Delta})] and $N_{\mathrm{tot}}$
[Eq.~(\ref{eq:FailProb&Delta})] with respect to the imperfection parameters
$p_{L}$ and $1-F$, while assuming $p_{I}=p_{M}=5\%$. Actually the choice of
$p_{I}$ and $p_{M}$ ($<10\%$) has negligible effect on the contours, since
they only modify $\varepsilon_{M}$ marginally. For initial fidelity
$F_{0}>0.95$, the contours of $\varepsilon_{E}$ are very close to vertical
lines; that is $\varepsilon_{E}$ is mostly limited by $p_{L}$ with an overhead
factor (about $10$) very insensitive to $F_{0}$. The contours of $N_{tot}$
indicate that the entanglement pumping needs about tens or hundreds of raw
Bell pairs to ensure a very high success probability.

Similarly, we may also numerically obtain the value $N_{\mathrm{tot}}$ from
Eq.~(\ref{eq:AIF2Delta}). The contour plot of $N_{\mathrm{tot}}$ with respect
to the imperfection parameters $p_{L}$ and $1-F$ is also shown in
Fig.~\ref{fig:Contours4PS} (c,f). We compare $N_{\mathrm{tot}}$s obtained from
two estimates (total error probability [Eq.~(\ref{eq:TEP2Delta})] and average
infidelity [Eq.~(\ref{eq:AIF2Delta})]). As we expected, the $N_{\mathrm{tot}}$
obtained from total error probability is approximately $1.2\sim2$ time larger
than the $N_{\mathrm{tot}}$ obtained from average infidelity, since the former
is a more conservative estimate and requires more unpurified Bell pairs.
Nevertheless, the difference is small and can be easily accounted by a
prefactor of order unity. For clarity, in the rest of the paper we will use
the $N_{\mathrm{tot}}$ estimated by using total error probability, and
sometimes quote the values estimated by using average infidelity.

\subsection{Entanglement pumping with non-post-selective (NPS) scheme}

We now consider another entanglement pumping protocol, proposed by Campbell
\cite{Campbell07}. The entanglement pumping scheme we have considered so far
is post-selective (PS); that is we discard the Bell pair if one step of
entanglement pumping is not successful. However, the Bell pair may still be
highly entangled even if the entanglement pumping failed at some intermediate
step. The non-post-selective (NPS) entanglement pumping scheme
\cite{Campbell07} keeps track of the evolution of the density matrix of the
Bell pair after each step of pumping. The NPS scheme avoids the inefficient
restart (i.e., discarding intermediately purified Bell pairs), and it may
reduce the time overhead, especially when the unpurified Bell pairs have
relatively low fidelity ($F<0.9$). In Ref. \cite{Campbell07}, the NPS pumping
is discussed in the context of generating a graph state.

We now describe how to use the NPS pumping scheme to generate purified Bell
pairs. To simplify the discussion, we first assume that the errors from local
measurements and operations are negligible. This assumption enables us to
establish a connection between the Markov chain model and the NPS pumping scheme.

\begin{figure}[tb]
\begin{center}
\includegraphics[
height=4.5 cm
]{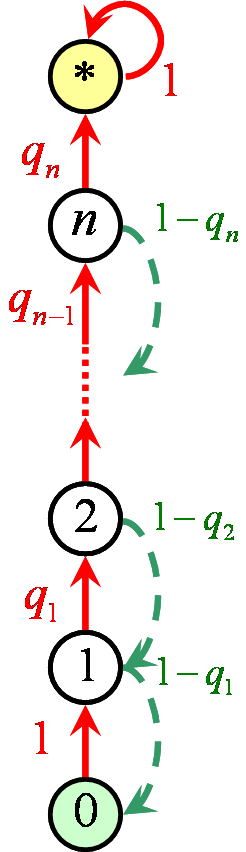}
\end{center}
\caption[fig:MarkovChain3]{(Color online) Markov chain model for one-level
entanglement pumping with non-post-selective (NPS) scheme. The key difference
from the previous Markov chain model with post-selective (PS) pumping scheme
(see Fig.~\ref{fig:MarkovChain1}) is that here the transition for unsuccessful
pumping reduces the chain label (score) by $1$, while in the previous model
the transition for unsuccessful pumping goes back to state "$0$" (restart of
the entire pumping scheme).}%
\label{fig:MarkovChain3}%
\end{figure}

Suppose the unpurified Bell pairs have only phase errors, then one level of
entanglement pumping is sufficient. For this error model, one can show that a
failed attempt produces an EPR pair with a density matrix identical to the one
in the previous step \cite{Campbell07}. One may introduce an accumulated score
associated with entanglement pumping. The score increases by one unit for each
attempt of successful pumping, and decrease by one unit for an attempt of
unsuccessful pumping. The score for no Bell pair is $0$, and for one
unpurified Bell pair it is $1$. The score exactly corresponds to the state
label of the Markov chain (see Fig.~\ref{fig:MarkovChain3}). After each
attempt of pumping, the score changes by $\pm1$. If the score drops to $0$
(i.e. no Bell pair left), it gets back to $1$ in the next attempt (i.e.,
creating a new unpurified Bell pair). The pumping procedure continues, until
the score reaches $n+1$ (i.e., the final state "$\ast$" in the Markov chain).
The key different from the previous Markov chain for post-selective pumping
scheme (see Fig.~\ref{fig:MarkovChain1}) is that here the score decrease by
$1$ for unsuccessful pumping rather than restart from $0$. This modification
increases the success probability of the robust entanglement generation.

\begin{figure}[tb]
\begin{center}
\includegraphics[
width=8.7 cm
]{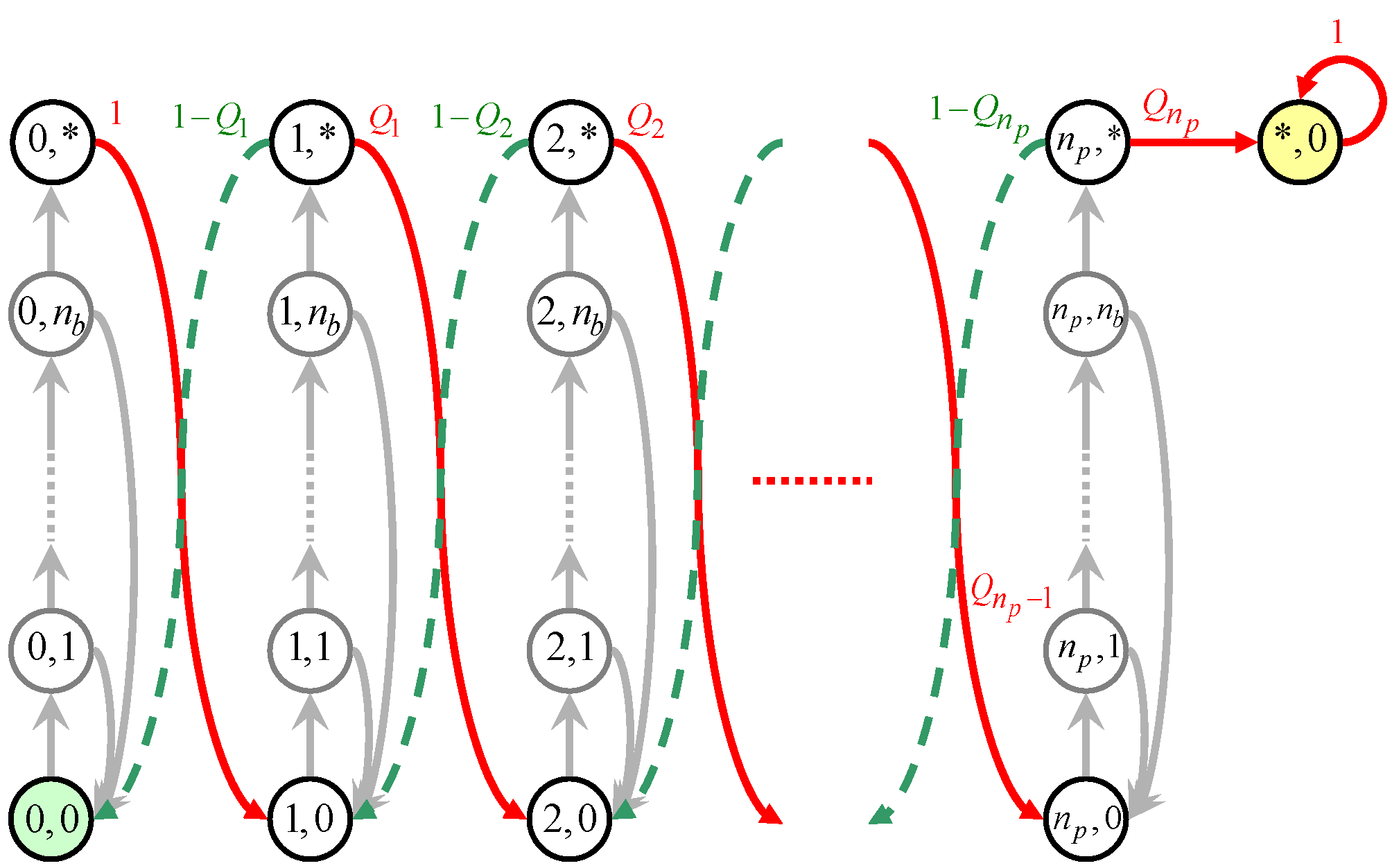}
\end{center}
\caption[fig:MarkovChain4]{(Color online) Markov chain model for two-level
entanglement pumping with NPS scheme. We still use PS entanglement pumping
scheme at the first level to have minimal accumulation of phase errors. Only
at the second level, does the NPS\ scheme work more efficiently than the
original PS scheme (see Fig.~\ref{fig:MarkovChain2}). }%
\label{fig:MarkovChain4}%
\end{figure}

When the unpurified Bell pairs have both bit-flip and phase errors (e.g.,
depolarizing error), we may use the bit-phase two-level pumping scheme (see
Sec. \ref{sec:two-level pumping scheme}), which purifies the bit error at the
first level and then the phase error at the second level. Since the phase
error is not purified at the first level, it accumulates after each attempt of
pumping. Therefore, it is better to use PS entanglement pumping scheme at the
first level to have minimal accumulation of phase errors. At the second level,
the NPS\ scheme works more efficiently than the PS scheme. The Markov chain
circuit for such mixed PS-NPS pumping schemes is shown in
Fig.~\ref{fig:MarkovChain4}.

In practice, the error probability for the local operations is always finite.
Then our simple Markov chain model only provides an approximate description
for the real process. The approximation comes from the fact that the score is
now insufficient to specify the density matrix for intermediate Bell pairs, in
the presence of local operational errors. In order to obtain the density
matrix for the intermediate state, we need to have the entire list of all
previous pumping outcomes. Nevertheless, when the local operational errors are
small compared to the infidelity of the intermediate Bell pairs, the Markov
chain model still provides an (optimistic) estimate for the total error
probability and the average fidelity.

\begin{figure}[tb]
\begin{center}
\includegraphics[
width=5.5 cm
]{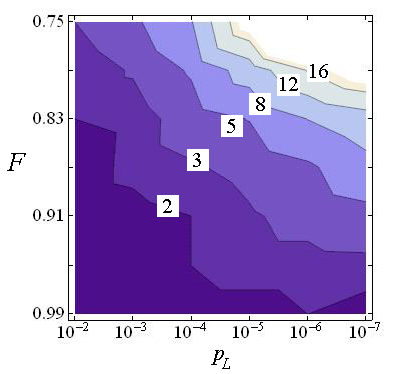}
\end{center}
\caption[fig:ContourPSNPS]{(Color online) The contours for the ratio between
$N_{\mathrm{tot}}$ associated with the post-selective (PS) scheme and
$N_{\mathrm{tot}}$ associated with the non-post-selective (NPS) scheme, as a
function of $p_{L}$ and $F$. For both schemes, we use the same formula
[Eq.~(\ref{eq:AIF2Delta})], but different Markov chain models
(Fig.~\ref{fig:MarkovChain1} and \ref{fig:MarkovChain3}). The improvement from
the NPS scheme becomes significant (more than a factor of $3$), for $F<0.9$
and $p_{L}<10^{-4}$. }%
\label{fig:ContourPSNPS}%
\end{figure}

We now compare the $N_{\mathrm{tot}}$s associated with the PS and NPS schemes.
The contours of the ratio between the two $N_{\mathrm{tot}}$s is plotted as a
function of $p_{L}$ and $F$ in Fig.~\ref{fig:ContourPSNPS}. As pointed out in
Ref. \cite{Campbell07}, there is a significant improvement by using the NPS
scheme (more than a factor of $3$), for $F<0.9$ and $p_{L}<10^{-4}$.

\section{Mapping to Deterministic Model\label{Mapping to Deterministic Model}}

In this section, we will map our stochastic, hybrid, and distributed quantum
computation scheme to a deterministic computation model, which is
characterized by two parameters --- the clock cycle and the effective error
probability. We will show that even when the underlying operations such as the
entanglement generation are non-deterministic, our approach still maintains
reasonable fast clock cycle time and sufficiently low effective error
probability. We will associate our discussion with achievable experimental
parameters, consider the constraint set by the finite memory lifetime, and
determine the achievable performance of hybrid distributed quantum computation.

\subsection{Time and error in the theoretical model}

All the previous discussions can be summarized in terms of the \emph{clock
cycle time}%
\begin{equation}
t_{C}=\tilde{t}_{E}+2t_{L}+\tilde{t}_{M}\approx\tilde{t}_{E}\ , \label{eq:tc}%
\end{equation}
and the \emph{effective error probability}
\begin{equation}
\gamma=\varepsilon_{E}+2p_{L}+2\varepsilon_{M}, \label{eq:gamma}%
\end{equation}
for a general coupling gate between two registers.

We now provide an estimate of the clock cycle time based on realistic
parameters. The time for optical initialization/measurement is
\begin{equation}
t_{I}=t_{M}\approx\frac{\ln p_{M}}{\ln\left(  1-\eta\right)  }\frac{\tau}{C},
\label{eq:ti0}%
\end{equation}
with a photon collection/detection efficiency $\eta$, vacuum radiative
lifetime $\tau$, and the cooperativity (Purcell) factor $C$ for
cavity-enhanced radiative decay \cite{Michler00,Purcell46}. Eq.~(\ref{eq:ti0})
is obtained from the estimate for the measurement error probability
$p_{M}\approx\left(  1-\eta\right)  ^{N_{\mathrm{photon}}}$ with
$N_{\mathrm{photon}}\approx t_{M}/\left(  \tau/C\right)  $. We assume that the
entanglement is generated based on detection of two photons
\cite{Duan03,Simon03}, which takes time
\begin{equation}
t_{E}\approx\left(  t_{I}+\tau/C\right)  /\eta^{2}. \label{eq:te0}%
\end{equation}
Generally entanglement fidelity is higher for the two-photon schemes than
one-photon schemes \cite{Childress05}. In addition, some two-photon schemes
have intrinsic purification against bit-flip errors \cite{Barrett05}. The time
for robust measurement is given in Eq.~(\ref{eq:tm}), and the total time for
robust entanglement generation is given in Eq.~(\ref{eq:te}).

Combining Eqs. (\ref{eq:tc}), (\ref{eq:ti0}), (\ref{eq:te0}), (\ref{eq:tm})
and (\ref{eq:te}), we obtain the clock cycle time (in units of the local
operation time) as a function of other parameters%
\begin{equation}
\frac{t_{C}}{t_{L}}=f\left[  \frac{\tau}{t_{L}C},p_{M},\eta,m,N_{\mathrm{tot}%
}\right]  .
\end{equation}
Meanwhile, we may obtain the relation $m=m\left[  p_{L},p_{I},p_{M}\right]  $
by minimizing $\varepsilon_{M}$ with Eq.~(\ref{eq:em}), and find the relation
$N_{\mathrm{tot}}=N_{\mathrm{tot}}\left[  p_{L},F,\varepsilon_{M}\right]
=N_{\mathrm{tot}}\left[  p_{L},F,2\Delta_{\min}\left[  p_{L},p_{I}%
,p_{M}\right]  \right]  $ using Eqs. (\ref{eq:DeltaMin} and \ref{eq:TEP2Delta}%
). Therefore, we have%
\begin{equation}
\frac{t_{C}}{t_{L}}=f\left[  \frac{\tau}{t_{L}C},p_{M},\eta,p_{L},F\right]  .
\end{equation}
The dimensionless parameter is the ratio between the times of emitting a
single photon and performing a local unitary operation. For systems such as
ion-traps and NV\ centers, this ratio is usually much less than unity ($<0.01$).

\begin{figure}[tb]
\begin{center}
\includegraphics[
width=8.7 cm
]{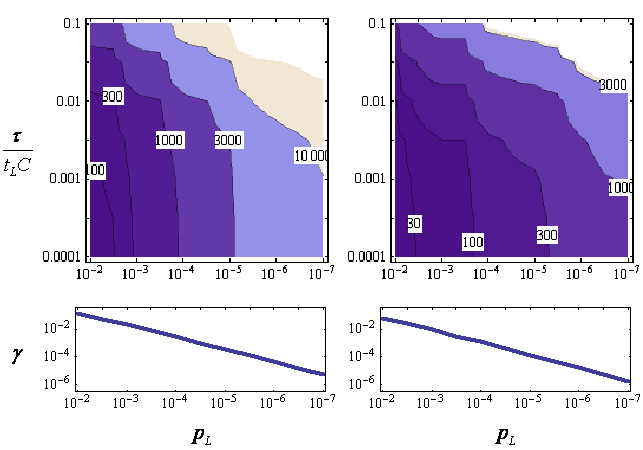}
\end{center}
\caption[fig:Time\&Fid1]{(Color online) Plots of clock cycle time $t_{C}$ and
effective error probability $\gamma$, for two-level pumping against
depolarizing errors. Upper plots: contours of the normalized clock cycle time
$t_{C}/t_{L}$ (upper) as a function of $p_{L}$ and $\frac{\tau}{t_{L}C}$ (the
normalized effective radiative lifetime). Lower plots: the effective error
probability $\gamma$ as a function of $p_{L}$. We assume $1-F=p_{I}=p_{M}=5\%$
(left) and $1\%$ (right), and $\eta=0.2$. }%
\label{fig:Time&Fid1}%
\end{figure}

Similarly, we can obtain the effective error probability in terms of
imperfection parameters%
\begin{equation}
\gamma=g\left[  p_{L},p_{M},F\right]  ,
\end{equation}
by combining Eqs. (\ref{eq:em}), (\ref{eq:gamma}) and (\ref{eq:TEP2Delta}).

\begin{figure}[tb]
\begin{center}
\includegraphics[
width=8.7 cm
]{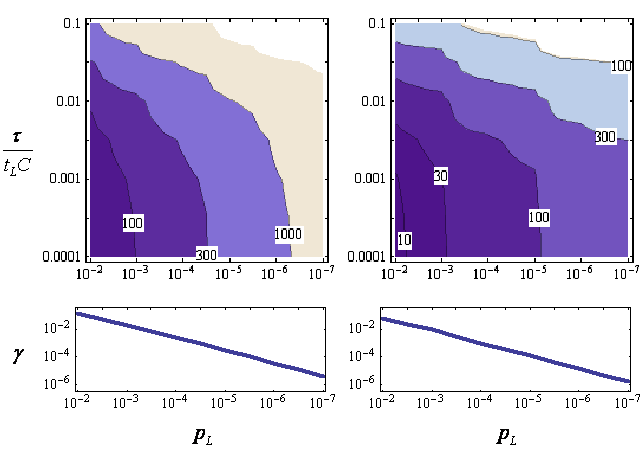}
\end{center}
\caption[fig:Time\&Fid2]{(Color online) Plots of clock cycle time $t_{C}$ and
effective error probability $\gamma$, for one-level pumping against dephasing
error. Upper plots: contours of the normalized clock cycle time $t_{C}/t_{L}$
(upper) as a function of $p_{L}$ and $\frac{\tau}{t_{L}C}$ (the normalized
effective radiative lifetime). Lower plots: the effective error probability
$\gamma$ as a function of $p_{L}$. We assume $1-F=p_{I}=p_{M}=5\%$ (left) and
$1\%$ (right). And $\eta=0.2$.}%
\label{fig:Time&Fid2}%
\end{figure}

In Fig.~\ref{fig:Time&Fid1}, we plot the clock cycle time $t_{C}$ and
effective error probability $\gamma$, for two-level pumping against
depolarizing error. Assuming $\eta=0.2$, we consider the two choices of
parameters $1-F=p_{I}=p_{M}=5\%$ (left) and $1\%$ (right). For each case, we
plot the contours of the normalized clock cycle time $t_{C}/t_{L}$ as a
function of $p_{L}$ and $\frac{\tau}{t_{L}C}$, and the effective error
probability $\gamma$ as a function of $p_{L}$. The clock cycle time can be
reduced by having a fast radiative decay rate $\tau/C$, which can be
facilitated by having a large cooperativity factor $C$. The reduction of the
clock cycle time stops once this ratio is below certain value, approximately
$0.003$ (left) and $0.001$ (right), where local gate operation becomes the
dominant time consuming step. Similarly, we plot the clock cycle time $t_{C}$
and effective error probability $\gamma$, for one-level pumping against
dephasing error in Fig.~\ref{fig:Time&Fid2}.

In the limit of negligible radiative decay time, we obtain the lower bound for
the normalized clock cycle time
\begin{equation}
\lim_{\frac{\tau}{t_{L}C}\rightarrow0}t_{C}/t_{L}=\left\{
\begin{tabular}
[c]{cc}%
$N_{\mathrm{tot}}$ & for $m=0$\\
$\left(  2m+2\right)  N_{\mathrm{tot}}$ & for $m\geq1$%
\end{tabular}
\ \right.  , \label{eq:tctl}%
\end{equation}
where for $m\geq1$ there is a time overhead $2m+2$ associated with local
operation and robust measurement; while there is no such overhead for $m=0$.

\subsection{Estimated numbers for experimental setups}

Suppose the parameters are $\left(  t_{L},\tau,\eta,C\right)  =\left(
0.1\text{ }\mu\text{s, }10\text{ ns, }0.2\text{, }10\right)  $
\cite{Garcia-Ripoll03,Steinmetz06,Keller04} and $\left(  1-F,p_{I},p_{M}%
,p_{L},\varepsilon_{M}\right)  =\left(  5\%,5\%,5\%,10^{-4},8\times
10^{-4}\right)  $ for our quantum registers (based on ion-traps or NV
centers). For depolarizing errors, two-level pumping can achieve $\left(
t_{C}\text{,}\gamma\right)  =\left(  200\text{ }\mu\text{s, }2.7\times
10^{-3}\right)  $. For some entanglement generation schemes
\cite{Duan03,Simon03,Barrett05} in principle only dephasing error exists,
because they have intrinsic purification against bit-flip errors. If all
bit-flip errors are suppressed, then one-level pumping is sufficient and
$\left(  t_{C}\text{,}\gamma\right)  =\left(  42\text{ }\mu\text{s, }%
2.2\times10^{-3}\right)  $\footnote[100]{For depolarizing or dephasing errors,
the estimated $t_{C}$ is approximately $1.3$ times less by using the average
infidelity.}.
In Table~\ref{table1}, we have listed $\left(  t_{C}\text{,}\gamma\right)  $
for parameters $1-F=p_{M}=p_{I}=5\%$ or $1\%$, and $p_{L}=10^{-3}$, $10^{-4}$,
$10^{-5}$, or $10^{-6}$. As expected, we find that $t_{C}$ gets longer, if the
fidelity $F$ is lower and/or the error probability ($p_{M}$ or $p_{I}$) is
higher; $t_{C}$ significantly reduces if the error for the unpurified Bell
pairs changes from depolarizing error to dephasing error.

\begin{table*}[tb]
\centering%
\begin{tabular}
[c]{p{2.3cm}ccp{0.4cm}ccp{0.5cm}ccp{0.4cm}cc}\hline\hline
\multicolumn{1}{c}{} & \multicolumn{5}{c}{Depolarizing} &  &
\multicolumn{5}{c}{Dephasing}\\\cline{2-12}%
\multicolumn{1}{c}{} & \multicolumn{2}{c}{$F=0.95$} &  &
\multicolumn{2}{c}{$F=0.99$} &  & \multicolumn{2}{c}{$F=0.95$} &  &
\multicolumn{2}{c}{$F=0.99$}\\\cline{2-12}
& $t_{C}$($\mu$s) & $\gamma$ &  & $t_{C}$($\mu$s) & $\gamma$ &  & $t_{C}$%
($\mu$s) & $\gamma$ &  & $t_{C}$($\mu$s) & $\gamma$\\\hline
$p_{L}=10^{-3}$ & 65 & $1.9\times10^{-2}$ &  & 19 & $9.1\times10^{-3}$ &  &
20 & $1.7\times10^{-2}$ &  & 8 & $8.7\times10^{-3}$\\\hline
$p_{L}=10^{-4}$ & 200 & $2.7\times10^{-3}$ &  & 49 & $1.2\times10^{-3}$ &  &
42 & $2.2\times10^{-3}$ &  & 17 & $9.9\times10^{-4}$\\\hline
$p_{L}=10^{-5}$ & 387 & $3.5\times10^{-4}$ &  & 65 & $1.3\times10^{-4}$ &  &
80 & $2.8\times10^{-4}$ &  & 22 & $1.2\times10^{-4}$\\\hline
$p_{L}=10^{-6}$ & 997 & $4.5\times10^{-5}$ &  & 162 & $1.7\times10^{-5}$ &  &
140 & $3.4\times10^{-5}$ &  & 39 & $1.4\times10^{-5}$\\\hline\hline
\end{tabular}
\caption[table1]{We list the values of $t_{C}$ and $\gamma$ as a function of
$p_{L}$ (rows) and $F$ (columns) for depolarizing and dephasing errors of the
unpurified Bell pairs. We also assume $p_{M}=p_{I}=1-F$, and $\left(
t_{L},\tau,\eta,C\right)  =\left(  0.1\text{ }\mu\text{s, }10\text{ ns,
}0.2\text{, }10\right)  $. Note that $t_{C}$ estimated by using average
infidelity is approximately $1.3\sim1.6$ times less than the numbers listed
here.}%
\label{table1}%
\end{table*}

We remark that $t_{C}$ should be much shorter than the memory time of the
storage qubit, $t_{mem}$. Because the memory error probability for each clock
cycle is approximately $t_{C}/t_{mem}$, which should be small (say $10^{-4}$)
in order to achieve fault-tolerant quantum computation. This is indeed the
case for both trapped ions (where $t_{mem}\sim10$ s has been demonstrated
\cite{Langer05,Haffner05b}), as well as for proximal nuclear spins of
NV\ centers (where $t_{mem}$ approaching a second can be inferred
\cite{Dutt07}). So far, we have justified the feasibility of the hybrid
distributed quantum computation scheme. In the next subsection, we will
provide a criterion for hybrid distributed quantum computation.

\subsection{Constraints from finite memory life time}

Above we have mostly ignored the effect of finite memory time, and with the
various sequences of purification of imperfections the final fidelity of the
operations have then been limited only be the local operation. All of these
purifications, however, increase the time of the operations and eventually the
system may become limited by the finite life time of the memory. In this
subsection we shall evaluate this constraint set by the finite memory lifetime.

\begin{figure*}[tb]
\begin{center}
\includegraphics[
width=13 cm
]{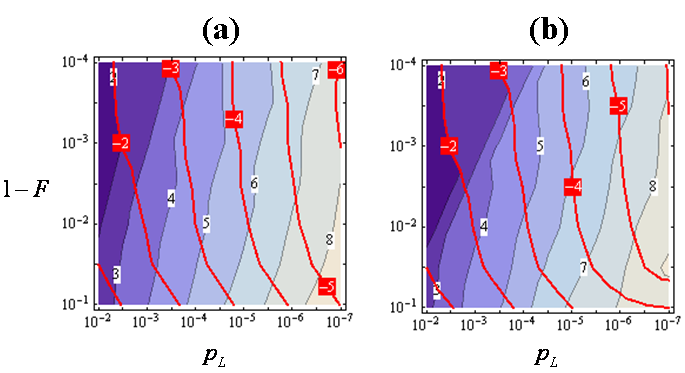}
\end{center}
\caption[fig:Contours4Tmem]{(Color online) Contours for $t_{mem}/t_{L}$
(shaded contours) and $\gamma$ (red contours), as a function of $p_{L}$ and
$1-F$. The label for the contour values are in logarithmic scale with base 10.
We consider both situations of (a) depolarizing error and (b) dephasing error.
The other parameters are $p_{M}=p_{I}=1-F$ and $\tau/C=0$.}%
\label{fig:Contours4Tmem}%
\end{figure*}

To simplify the discussion we assume that we have a very short radiative
lifetime $\tau$ or that we are able to achieve a very large Purcell factor so
that $\tau/C$ becomes negligible. All the time scales are then proportional to
the local gate time $t_{L}$. With a finite memory time, i.e., some fixed
$t_{mem}/t_{L}$, there is a limit to have many operations we can do before we
are limited by the memory error. To get an estimate for this limit we assume
that the ideal number of operations is roughly given by the point, where the
memory error probability is the same as the effective error probability for
the non-local coupling gate:%
\begin{equation}
t_{C}/t_{mem}=\gamma.
\end{equation}
Then according to Eq.~(\ref{eq:tctl}), we have%
\begin{equation}
\frac{t_{mem}}{t_{L}}=\frac{t_{mem}}{t_{C}}\frac{t_{C}}{t_{L}}=\gamma
^{-1}\left(  2m+2-\delta_{m,0}\right)  N_{\mathrm{tot}},
\end{equation}
where the variables $\left\{  \gamma,m,N_{\mathrm{tot}}\right\}  $ are all
determined by the imperfection parameters $\left\{  1-F,p_{M},p_{I}%
,p_{L}\right\}  $. We further reduce the imperfection parameters by assuming
$1-F=p_{M}=p_{I}$, and get the contour plot of $t_{mem}/t_{L}$ in terms of the
imperfection parameter $p_{L}$ and $1-F$ in Fig.~\ref{fig:Contours4Tmem}. In
the plot, we consider both the situation of depolarizing or dephasing error
during entanglement generation.

For given $t_{mem}/t_{L}$, we may use Fig.~\ref{fig:Contours4Tmem} to find the
valid region in the parameter space of $p_{L}$ and $1-F$, and then identify
the achievable effective error probability $\gamma$. For example, with
ion-trap systems it may be possible to achieve $t_{mem}/t_{L}\sim10^{8}$
\cite{Garcia-Ripoll03,Langer05,Haffner05b}, and the region left of the shaded
contour line ($\log_{10}t_{mem}/t_{L}=8$) can then be accessed, which enables
us to obtain a wide range effective error probability $\gamma$ depending on
the practical values of $p_{L}$ and $1-F$. For NV centers, it should be
feasible to achieve $t_{mem}/t_{L}\sim10^{7}$ by having $t_{mem}\approx10$ s
and $t_{L}\sim10^{-6}$ s \cite{Dutt07}; the region on the left side of the
shaded contour line ($\log_{10}t_{mem}/t_{L}=7$) still covers a large portion
of the parameter space. For a given experiment situation with a finite memory
time as well as other imperfections, we can thus use
Fig.~\ref{fig:Contours4Tmem} to determine the achievable performance of hybrid
distributed quantum computation.

\section{Approaches to Fault Tolerance\label{Fault Tolerance}}

The entanglement based approach discussed in this paper provides a method to
make gates between any quantum registers and this can be used to implement
arbitrary quantum circuits, once the errors in the gates are sufficiently
small. The errors can be further suppressed by using quantum error correction.
For example, as shown in Table~\ref{table1}, $\left(  p_{L},F\right)  =\left(
10^{-4},0.95\right)  $ can achieve $\gamma\approx2.7\times10^{-3}$, well below
the $1\%$ threshold for fault tolerant computation based on approaches such as
the $C_{4}/C_{6}$ code~\cite{Knill05} or 2D toric codes~\cite{Raussendorf07};
$\left(  p_{L},F\right)  =\left(  10^{-6},0.99\right)  $ can achieve
$\gamma\approx1.7\times10^{-5}$, which allows efficient codes such as the BCH
[[127,43,13]] code to be used without concatenation.

Following Ref.~\cite{Steane03} we estimate $20$ registers per logical qubit to
be necessary for a calculation involving $K=10^{4}$ logical qubits and
$Q=10^{6}$ logical operations, assuming the memory failure rate and effective
error probability are $t_{C}/t_{mem}\approx\gamma\approx1.7\times10^{-5}$
(e.g., achieved by $t_{mem}\approx10$ s, $t_{C}\approx162$~$\mu$s). (This
estimate is based on Fig.~10b of Ref.~\cite{Steane03}.) Assuming that error
correction is applied after each logical operation, and that logical operation
and following recovery take approximately $2-16$ clock cycles depending on the
type of operation and the coding scheme (see section II.A of
Ref.~\cite{Steane03}), the total running time of this computation would then
be approximately $400-3000$ s.

We remark that one important property of distributed quantum computation is
that the measurement time is relatively fast compared with the non-local
coupling gate, because the measurement does not rely on the time-consuming
processes of entanglement generation and purification while the non-local
coupling gate does. This property is different from the conventional model of
quantum computation, where the measurement is usually a slow process that
induces extra overhead in both time and physical resources \cite{Steane03}.
Thus, instead of reconciling slow measurements \cite{Divincenzo07}, it might
also be interesting to study possible improvement using fast measurements for
fault-tolerant quantum computation.

\begin{figure}[tb]
\begin{center}
\includegraphics[
width=8.7 cm
]{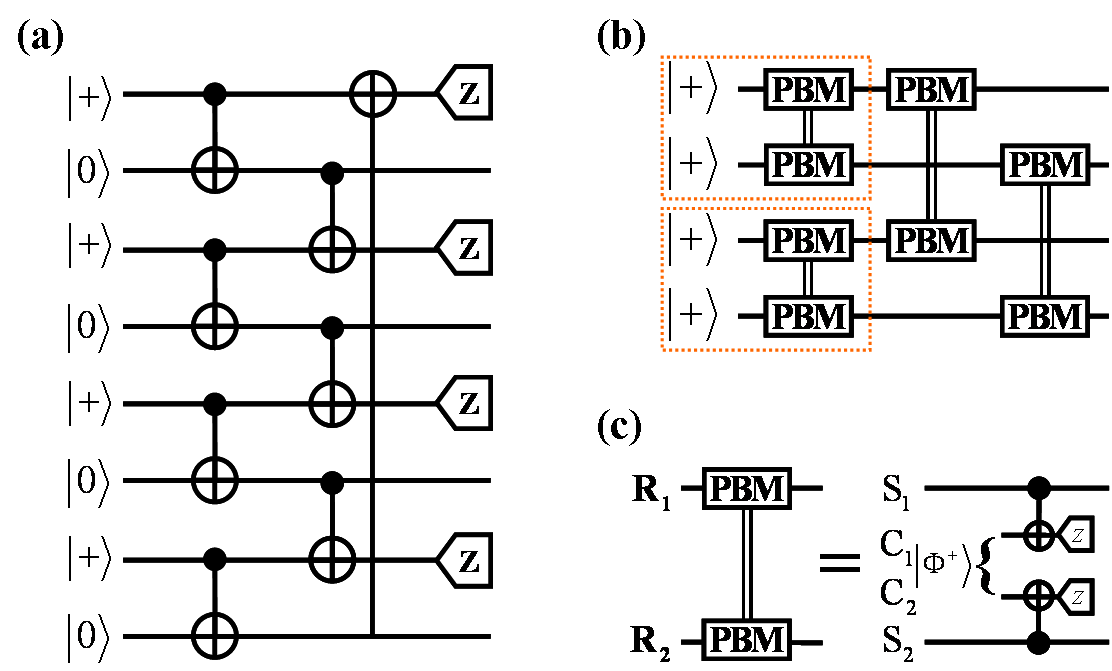}
\end{center}
\caption[fig:GHZ]{(Color online) Circuits for fault-tolerant preparation of
GHZ\ state. (a) Conventional circuit uses $8$ quantum registers to generate a
$4$-qubit GHZ state. The state $\left\vert +\right\rangle \ $is $\left(
\left\vert 0\right\rangle +\left\vert 1\right\rangle \right)  /\sqrt{2}$. (b)
New circuit requires only $4$ registers, by using partial Bell measurement
(PBM) between registers, detailed in (c).}%
\label{fig:GHZ}%
\end{figure}

The above estimates have been performed assuming that our hybrid register
based approach is mapped directly to the standard circuit model. In some
situations this may, however, not be the most advantageous way to proceed,
since the register architecture may allow for more efficient performance of
certain tasks. As a particular example, we now briefly discuss an alternative
new approach to fault-tolerant preparation of GHZ states (e.g., $\left\vert
00\cdots0\right\rangle +\left\vert 11\cdots1\right\rangle $), which are a
critical component both for syndrome extraction and construction of universal
gates in quantum error correcting codes \cite{NC00,Knill05}. This new approach
relies upon the observation that the EPR pairs from entanglement generation
can be used for \emph{deterministic} partial Bell measurement (PBM), which is
achieved by applying local coupling gates and projective measurements as shown
in Fig.~\ref{fig:GHZ}c. After the PBM, the two storage qubits are projected to
the subspace spanned by the Bell states $\left\vert \Phi^{\pm}\right\rangle
=\left(  \left\vert 00\right\rangle \pm\left\vert 11\right\rangle \right)
/\sqrt{2}$ if the measurement outcomes are the same, or they are projected to
the subspace spanned by the Bell states $\left\vert \Psi^{\pm}\right\rangle
=\left(  \left\vert 01\right\rangle \pm\left\vert 10\right\rangle \right)
/\sqrt{2}$ if the measurement outcomes are different. For the latter case, we
may further flip one of the storage qubits, so that they are projected to the
subspace spanned by $\left\vert \Phi^{\pm}\right\rangle $. By using PBMs, we
can perform fault-tolerant preparation of GHZ state efficiently (up to single
qubit rotations) as detailed below.

Fault-tolerant state preparation requires that the probability to have errors
in more than one qubit in the prepared state is $O\left(  p^{2}\right)  $,
with the error probability for each input qubit or quantum gate being
$O\left(  p\right)  $; that is multiple errors only occur at the second or
higher orders. The regular circuit to prepare a four-qubit GHZ\ state
fault-tolerantly \cite{Knill05} is shown in Fig.~\ref{fig:GHZ}a. If this
circuit should be implemented with quantum registers, the CNOT gates in
Fig.~\ref{fig:GHZ}a should be created by using the circuit detailed in
Fig.~\ref{fig:CNOT}, and eight quantum registers would be required.

By using PBMs, however, only \emph{four} quantum registers are needed in order
to generate GHZ states fault-tolerantly as shown in Fig.~\ref{fig:GHZ}b. The
fault-tolerance comes from the last (redundant) PBM between the second and
fourth register (Fig.~\ref{fig:GHZ}b), which detects bit-errors from earlier
PBMs. The advantage of the PBMs is that it propagates neither bit- nor
phase-errors. The circuit of Fig.~\ref{fig:GHZ}c indicates that the only way
to propagate error from one input to another (say, $S_{1}$ to $S_{2}$) is via
some initial error in the entangled pair between $C_{1}$ and $C_{2}$. However,
for Bell states $\left\vert \Phi^{\pm}\right\rangle $, we have the following
identities
\begin{align}
X_{C_{2}}\left\vert \Phi^{\pm}\right\rangle _{C_{1},C_{2}}  &  =\pm X_{C_{1}%
}\left\vert \Phi^{\pm}\right\rangle _{C_{1},C_{2}}\\
Z_{C_{2}}\left\vert \Phi^{\pm}\right\rangle _{C_{1},C_{2}}  &  =Z_{C_{1}%
}\left\vert \Phi^{\pm}\right\rangle _{C_{1},C_{2}}.
\end{align}
Similar identities also exist for Bells states $\left\vert \Psi^{\pm
}\right\rangle $. Suppose $S_{1}$ has an error, because of the above
identities, we can always treat the imperfection of the Bell pair $\left\vert
\Phi^{+}\right\rangle _{C_{1},C_{2}}$ as an error in $C_{1}$ (the qubit from
the same register as $S_{1}$). Therefore, only the first register has errors
and they never propagate to the second one. (PBM may induce errors to
unconnected but entangled qubits.)

\begin{figure}[tb]
\begin{center}
\includegraphics[
width=7 cm
]{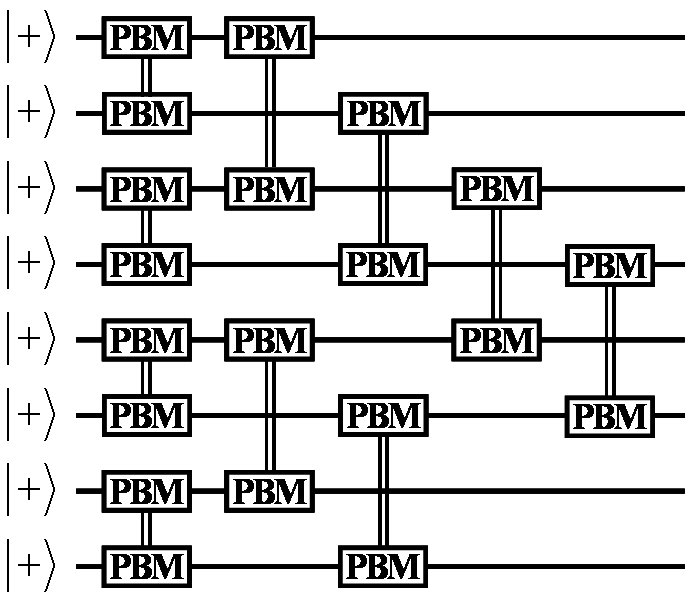}
\end{center}
\caption[fig:GHZ2]{Circuit for fault-tolerant preparation of the $8$-qubit GHZ
state with only $8$ quantum registers.}%
\label{fig:GHZ2}%
\end{figure}

The present scheme may be expanded to larger numbers of qubits and generally
we may fault-tolerantly prepare $2^{n}$-qubit GHZ state with only $2^{n}$
quantum registers, by recursively using Fig.~\ref{fig:GHZ}b with the two
dashed boxes replaced by two $2^{n-1}$-qubit GHZ states. The circuit for
fault-tolerant preparation of the $8$-qubit GHZ state is shown in
Fig.~\ref{fig:GHZ2}. Note that we can perform PBMs acting on different
registers in parallel. Suppose each PBM takes one clock cycle, the preparation
time is $2$ (clock cycles) for a 4-qubit GHZ\ state shown in
Fig.~\ref{fig:GHZ}b. The two PBMs in the orange boxes are performed in the
first clock cycle, and the rest for the second clock cycle. Generally, for a
$2^{n}$-qubit GHZ state with $n\geq3$ (see discussion in Appendix \ref{GHZ}),
the preparation time is only $3$ (clock cycles), and the error probability for
each register is only approximately $3p/2$. Therefore, the PBM-based scheme
for fault-tolerant preparation of the GHZ state is efficient in both time and physical-resources.

\section{Conclusion\label{Conclusion}}

In conclusion, we have proposed an efficient register-based, hybrid quantum
computation scheme. Our scheme requires only five qubits (or less) per
register, and it is robust against various kinds of imperfections, including
imperfect initialization/measurement and low fidelity entanglement generation.
We presented a Markov chain model to analyze the time overhead associated with
the robust operations of measurement and entanglement generation. We found
reasonable time overhead and considered practical implementation of quantum
registers with ion traps or NV centers. We also provided an
example using partial Bell measurement to prepare GHZ states for
fault-tolerant quantum computation. It might be possible to further facilitate
fault-tolerant quantum computation with systematic optimization using dynamic
programming \cite{JTKL07}.

\section*{Acknowledgements}

The authors wish to thank Gurudev Dutt, Lily Childress, Paola Cappellaro, Earl Campbell,
Wolfgang D\"{u}r, Phillip Hemmer, and Charles Marcus. This work is supported by NSF, DTO,
ARO-MURI, the Packard Foundations, Pappalardo Fellowship, and the Danish
Natural Science Research Council.

\appendix

\section{Bit-phase two-level pumping
scheme\label{Bit-phase two-level pumping scheme}}

\begin{figure}[tb]
\begin{center}
\includegraphics[
width=7 cm
]{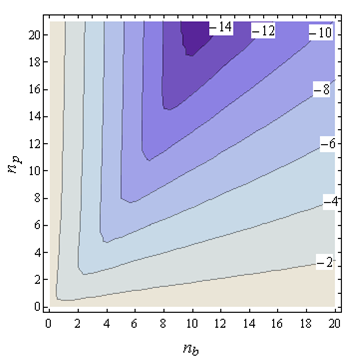}
\end{center}
\caption[fig:Appendix]{(Color online) The contours of the infidelity
$\varepsilon_{E,\text{\textrm{infid}}}^{\left(  n_{b},n_{p}\right)  }$ as a
function of $n_{b}$ and $n_{p}$. We use $\log_{10}\varepsilon
_{E,\text{\textrm{infid}}}^{\left(  n_{b},n_{p}\right)  }$ to label the
contours. We assume a depolarizing error with initial fidelity $F=0.95$, and
$\tilde{\varepsilon}_{M}=p_{L}=0$. The final infidelity can be arbitrarily
small for sufficiently large $n_{b}$ and $n_{p}$. This indicates that the
bit-phase two-level entanglement pumping scheme can create pairs with
arbitrarily high fidelity.}%
\label{fig:Appendix}%
\end{figure}

In this Appendix we show that our bit-phase two-level entanglement pumping
scheme can create pairs with fidelity arbitrarily close to unity, if we have
perfect local operations. Numerical indication of this is shown in
Fig.~\ref{fig:Appendix}, and in the following we will provide a rigorous proof
to this claim.

We assume that the initial state is a mixed state that has only diagonal terms
in the Bell basis%
\begin{equation}
\rho=a\left\vert \Phi^{+}\right\rangle \left\langle \Phi^{+}\right\vert
+b\left\vert \Phi^{-}\right\rangle \left\langle \Phi^{-}\right\vert
+c\left\vert \Psi^{+}\right\rangle \left\langle \Psi^{+}\right\vert
+d\left\vert \Psi^{-}\right\rangle \left\langle \Psi^{-}\right\vert ,
\label{eq:rou}%
\end{equation}
where $\left\vert \Phi^{\pm}\right\rangle =\left(  \left\vert 00\right\rangle
\pm\left\vert 11\right\rangle \right)  /\sqrt{2}$, $\left\vert \Psi^{\pm
}\right\rangle =\left(  \left\vert 01\right\rangle \pm\left\vert
10\right\rangle \right)  /\sqrt{2}$, and the coefficients are non-negative and
sum to unity. (This assumption is only made to simplify the presentation. For
a general density matrix, only the diagonal elements given in
Eq.~(\ref{eq:rou}) are important \cite{Deutsch96}.) After the purification the
density matrix retains this form but with new coefficients. Therefore, we only
need four coefficients for each state using Bell basis, denoted as the
fidelity vector $\vec{F}=\left(  a,b,c,d\right)  $. We use the lower index to
keep track of the pumping steps, so that the fidelity vector for the
unpurified state is $\vec{F}_{0}=\left(  a_{0},b_{0},c_{0},d_{0}\right)  $,
and the vector for the purified state after $n$ steps of entanglement pumping
is $\vec{F}_{n}=\left(  a_{n},b_{n},c_{n},d_{n}\right)  $.

Suppose we use the $\vec{F}_{0}$ state to pump the state $\vec{F}_{n}$ against
bit-errors, the success probability is%
\begin{equation}
p_{n+1}=\left(  a_{0}+b_{0}\right)  \left(  a_{n}+b_{n}\right)  +\left(
c_{0}+d_{0}\right)  \left(  c_{n}+d_{n}\right)  ,
\end{equation}
and the fidelity vector is%
\begin{align}
&  \vec{F}_{n+1}\\
&  =\frac{1}{p_{n+1}}\left(  a_{0}a_{n}+b_{0}b_{n},a_{0}b_{n}+b_{0}a_{n}%
,c_{0}c_{n}+d_{0}d_{n},c_{0}d_{n}+d_{0}c_{n}\right)  ,
\end{align}
for perfect local operations (measurement and CNOT gate).

Similarly, for pumping against phase-errors, the success probability is%
\begin{equation}
p_{n+1}^{\prime}=\left(  a_{0}+c_{0}\right)  \left(  a_{n}+c_{n}\right)
+\left(  b_{0}+d_{0}\right)  \left(  b_{n}+d_{n}\right)  ,
\end{equation}
and the fidelity vector is%
\begin{align}
&  \vec{F}_{n+1}^{\prime}\\
&  =\frac{1}{p_{n+1}^{\prime}}\left(  a_{0}a_{n}+c_{0}c_{n},b_{0}b_{n}%
+d_{0}d_{n},a_{0}c_{n}+c_{0}a_{n},b_{0}d_{n}+d_{0}b_{n}\right)  .
\end{align}

In general, any state can be turned into a so-called Werner state with the
same fidelity, and as a worst case scenario we shall assume the unpurified
Bell state to be a Werner state:%
\begin{equation}
\vec{F}_{0}=\left(  F_{0},\frac{1-F_{0}}{3},\frac{1-F_{0}}{3},\frac{1-F_{0}%
}{3}\right)  .
\end{equation}
with $F_{0}>1/2$ to ensure that it contains distillable entanglement. For
convenience of later discussion, we rewrite $\vec{F}_{0}$ as%
\begin{equation}
\vec{F}_{0}=\left(  1-3\alpha/2,\alpha/2,\alpha/2,\alpha/2\right)
\end{equation}
where $\alpha=\frac{2}{3}\left(  1-F_{0}\right)  \,<\frac{1}{3}$.

\subsection{First level pumping}

For the first level of pumping against bit-errors, we may parameterize the
fidelity vector%
\begin{equation}
\vec{F}_{n}=\left(  a_{n},b_{n},c_{n},d_{n}\right)  =\left(  \frac{1}%
{2}+\delta_{n},\frac{1}{2}-\delta_{n}-2\eta_{n},\eta_{n},\eta_{n}\right)
\end{equation}
in terms of two variables $\delta_{n}$ and $\eta_{n}$, and obtain some bounds
for these variables. Since we are pumping against bit-errors, $\eta_{n}$
decreases with $n$%
\begin{equation}
\eta_{n+1}<\eta_{n}<\cdots<\eta_{0}=\alpha/2<1/6.
\end{equation}
The success probability for the $\left(  n+1\right)  $th step of pumping is%
\begin{align}
p_{n+1}  &  =\left(  1-\alpha\right)  \left(  1-2\eta_{n}\right)  +2\alpha
\eta_{n}\label{eq:psucc1st}\\
&  <1-\alpha-2\alpha\eta_{n}<1-\alpha,
\end{align}
and on the other hand we have%
\begin{equation}
p_{n+1}>\left(  1-\alpha\right)  \left(  1-2\eta_{n}\right)  >\frac{2}%
{3}\left(  1-\alpha\right)
\end{equation}
where the second inequality follows from $\eta_{n}<1/6$ \footnote[100]{Note
that the lower bound for $p_{n+1}$ can be arbitrarily close to $1-\alpha$ for
sufficiently large $n$. Here, for clarity we use the loose bound $\frac{2}%
{3}\left(  1-\alpha\right)  <p_{n+1}$, which later imposes a constraint
$\alpha<2/7$ (i.e. $F>4/7\approx0.57$). In principle, by choosing tighter
lower bounds for $p_{n+1}$, the following proof works for all $\alpha<1/3$.}.

The third element of the fidelity vector is%
\begin{equation}
\eta_{n+1}=c_{n+1}=\frac{c_{0}\eta_{n}+d_{0}\eta_{n}}{p_{n+1}}=\frac{\alpha
}{p_{n+1}}\eta_{n}.
\end{equation}
Since $p_{n+1}>\frac{2}{3}\left(  1-\alpha\right)  $, we have%
\begin{align}
\eta_{n+1}  &  <\frac{\alpha}{\frac{2}{3}\left(  1-\alpha\right)  }\eta
_{n}<\cdots<\left(  \frac{3\alpha}{2\left(  1-\alpha\right)  }\right)
^{n+1}c_{0}\label{eq:eta1st}\\
&  <\left(  \frac{3}{4}\right)  ^{n+1}c_{0}\nonumber
\end{align}
which indicates that $\eta_{n}$ approaches zeros exponentially with respect to
$n$. When $n$ is large enough, $\eta_{n}$ is negligible and $p_{n}%
\approx1-\alpha-O\left(  \eta_{n}\right)  $.

Similarly, we obtain the recursive relation for $\delta_{n}$.
\begin{align}
\frac{1}{2}+\delta_{n+1}  &  =a_{n+1}=\frac{1}{p_{n+1}}\left(  a_{0}%
a_{n}+b_{0}b_{n}\right) \\
&  =\frac{1}{2}\frac{1-\alpha-2\alpha\eta_{n}}{p_{n+1}}+\frac{1-2\alpha
}{p_{n+1}}\delta_{n}\\
&  >\frac{1}{2}+\frac{1-2\alpha}{1-\alpha}\delta_{n}.
\end{align}
Thus%
\begin{align}
\delta_{n+1}  &  >\frac{1-2\alpha}{1-\alpha}\delta_{n}>\cdots>\left(
\frac{1-2\alpha}{1-\alpha}\right)  ^{n+1}\delta_{0}\label{eq:delta1st}\\
&  =\left(  \frac{1-2\alpha}{1-\alpha}\right)  ^{n+1}\frac{1-3\alpha}%
{2}\nonumber
\end{align}
and $\delta_{0}=\frac{1-3\alpha}{2}$. We have therefore%
\begin{equation}
\frac{\eta_{n}}{\delta_{n}}<\left(  \frac{2-4\alpha}{3\alpha}\right)
^{n}\frac{\alpha}{1-3\alpha} \label{eq:eta/delta1st}%
\end{equation}

\subsection{Second level pumping}

After $n_{b}$ steps of first level pumping, we have the fidelity vector%
\begin{equation}
\vec{F}_{0}^{\prime}=\left(  \frac{1}{2}+\delta_{0}^{\prime},\frac{1}%
{2}-\delta_{0}^{\prime}-2\eta_{0}^{\prime},\eta_{0}^{\prime},\eta_{0}^{\prime
}\right)
\end{equation}
where $\delta_{0}^{\prime}=\delta_{n_{b}}>\left(  \frac{1-2\alpha}{1-\alpha
}\right)  ^{n+1}\delta_{0}$ and $\eta_{0}^{\prime}=\eta_{n_{b}}<\left(
\frac{3\alpha}{2\left(  1-\alpha\right)  }\right)  ^{n+1}c_{0}$.

The fidelity vector after $n_{p}$ steps of pumping against phase errors is%
\begin{equation}
\vec{F}_{n}^{\prime}=\left(  \frac{1}{2}+\delta_{n}^{\prime},\frac{1}%
{2}-\delta_{n}^{\prime}-\eta_{n}^{\prime}-d_{n},\eta_{n}^{\prime}%
,d_{n}\right)  ,
\end{equation}
where $\delta_{n}^{\prime}<1/2$ and $\eta_{n}^{\prime}<1/2$.

The success probability for the $\left(  n+1\right)  $th step is%
\begin{align}
p_{n+1}^{\prime}  &  =\left(  \frac{1}{2}+\delta_{0}^{\prime}+\eta_{0}%
^{\prime}\right)  \left(  \frac{1}{2}+\delta_{n}^{\prime}+\eta_{n}^{\prime
}\right) \nonumber\\
&  \quad+\left(  \frac{1}{2}-\delta_{0}^{\prime}-\eta_{0}^{\prime}\right)
\left(  \frac{1}{2}-\delta_{n}^{\prime}-\eta_{n}^{\prime}\right) \nonumber\\
&  =\frac{1}{2}+2\left(  \delta_{0}^{\prime}+\eta_{0}^{\prime}\right)  \left(
\delta_{n}^{\prime}+\eta_{n}^{\prime}\right) \label{eq:psucc2nd}\\
&  >\frac{1}{2}\nonumber
\end{align}

We now consider the elements of $\vec{F}_{n}^{\prime}$. On one hand, the
erroneous admixture of $\left\vert \Psi^{-}\right\rangle $ described by
$\left\{  d_{n}\right\}  $ keep decreasing with $n$, since $\left\vert
\Psi^{-}\right\rangle $ errors are also purified during the second level
pumping. On the other hand, the erroneous admixture of $\left\vert \Psi
^{+}\right\rangle $ described by $\left\{  \eta_{n}^{\prime}\right\}  $ may
increase with $n$, but it is upper-bounded by the following relation:%
\begin{align}
\eta_{n+1}^{\prime}  &  =\frac{a_{0}c_{n}+c_{0}a_{n}}{p_{n+1}^{\prime}%
}\nonumber\\
&  <2\left[  \left(  \frac{1}{2}+\delta_{0}^{\prime}\right)  \eta_{n}^{\prime
}+\eta_{0}^{\prime}\left(  \frac{1}{2}+\delta_{n}^{\prime}\right)  \right]
\nonumber\\
&  <\left(  1+2\delta_{0}^{\prime}\right)  \eta_{n}^{\prime}+2\eta_{0}%
^{\prime}\label{eq:eta2nd}\\
&  <\left(  1+2\delta_{0}^{\prime}\right)  ^{n+1}\eta_{0}^{\prime},\nonumber
\end{align}
and one can also show the lower bound for $\eta_{n+1}^{\prime}$ by induction:
\begin{equation}
\eta_{n+1}^{\prime}>a_{0}c_{n}+c_{0}a_{n}>\frac{1}{2}\left(  \eta_{n}^{\prime
}+\eta_{0}^{\prime}\right)  >\eta_{0}^{\prime}.
\end{equation}

However, $\delta_{n}^{\prime}$ approaches $1/2$ much faster:%
\begin{align}
\frac{1}{2}+\delta_{n+1}^{\prime}  &  =a_{n+1}=\frac{a_{0}a_{n}+c_{0}c_{n}%
}{p_{n+1}^{\prime}}\nonumber\\
&  =\frac{\left(  \frac{1}{2}+\delta_{0}^{\prime}\right)  \left(  \frac{1}%
{2}+\delta_{n}^{\prime}\right)  +\eta_{0}^{\prime}\eta_{n}^{\prime}}{\frac
{1}{2}+2\left(  \delta_{0}^{\prime}+\eta_{0}^{\prime}\right)  \left(
\delta_{n}^{\prime}+\eta_{n}^{\prime}\right)  }\label{eq:delta2nd}\\
&  =\frac{1}{2}+\frac{\delta_{n}^{\prime}+\delta_{0}^{\prime}-2\left(
\delta_{0}^{\prime}\eta_{n}^{\prime}+\delta_{n}^{\prime}\eta_{0}^{\prime}%
+\eta_{0}^{\prime}\eta_{n}^{\prime}\right)  }{1+4\delta_{0}^{\prime}\delta
_{n}^{\prime}+4\left(  \delta_{0}^{\prime}\eta_{n}^{\prime}+\delta_{n}%
^{\prime}\eta_{0}^{\prime}+\eta_{0}^{\prime}\eta_{n}^{\prime}\right)
}\nonumber
\end{align}
If \underline{$\frac{1}{2}>\delta_{n}^{\prime}>\delta_{0}^{\prime}>\eta
_{n}^{\prime}>\eta_{0}^{\prime}$}, we have
\begin{equation}
\delta_{0}^{\prime}\eta_{n}^{\prime}+\delta_{n}^{\prime}\eta_{0}^{\prime}%
+\eta_{0}^{\prime}\eta_{n}^{\prime}<\delta_{0}^{\prime}\eta_{n}^{\prime}%
+\eta_{0}^{\prime}<\delta_{0}^{\prime}\zeta,
\end{equation}
where we in the second inequality introduced a number $\zeta$ such that
\underline{$\delta_{0}^{\prime}\eta_{n}^{\prime}+\eta_{0}^{\prime}<\delta
_{0}^{\prime}\zeta$}. In the next subsection, we will show that we may choose
$\zeta$ small (i.e., \underline{$\zeta=2\varepsilon$}) such that%

\begin{align}
\delta_{n+1}^{\prime}  &  >\frac{\delta_{n}^{\prime}+\delta_{0}^{\prime
}\left(  1-2\zeta\right)  }{1+4\delta_{0}^{\prime}\delta_{n}^{\prime}\left(
1+\zeta\right)  }\\
\lambda_{n+1}^{\prime}  &  >\frac{\lambda_{n}^{\prime}+\lambda_{0}^{\prime
}\left(  1-2\zeta\right)  }{1+4\lambda_{0}^{\prime}\left(  1-2\zeta\right)
\lambda_{n}^{\prime}}\nonumber\\
&  >\lambda_{n}^{\prime}+\lambda_{0}^{\prime}\left(  1-2\zeta\right)
-2\lambda_{0}^{\prime}\left(  1-2\zeta\right)  \lambda_{n}^{\prime}%
\end{align}
where $\lambda_{n}^{\prime}=\sqrt{\frac{1+\zeta}{1-2\zeta}}\delta_{n}^{\prime
}\approx\left(  1+3\zeta/2\right)  \delta_{n}^{\prime}$ for $n=0,1,\cdots$,
and the third equality uses $\frac{x+y}{1+4xy}>x+y-2xy$ for $0<x,y<1/2$.
Finally, we have%
\begin{align}
1/2-\lambda_{n+1}^{\prime}  &  <\left(  1-2\lambda_{0}^{\prime}\left(
1-2\zeta\right)  \right)  \left(  1/2-\lambda_{n}^{\prime}\right) \nonumber\\
&  <\left(  1-2\lambda_{0}^{\prime}\left(  1-2\zeta\right)  \right)
^{n+1}\left(  1/2-\lambda_{0}^{\prime}\right)  \label{eq:lumda2nd}%
\end{align}

So far, we have introduced inequalities to bound elements of fidelity vectors
at different stages of pumping. In the next subsection, we will use these
constraints to show that we are able to achieve fidelity arbitrarily close to
unity by carefully choosing the numbers of pumping steps $\left(  n_{b}%
,n_{p}\right)  $ for the bit-phase two-level pumping scheme.

\subsection{$\varepsilon-N$ argument}

For $\forall\varepsilon>0$ and $\alpha<2/7$, we may choose $n_{b}$
\begin{equation}
\fbox{$n_{b}\geq\max\left\{  \frac{\ln\varepsilon-\ln\frac{\alpha}{1-3\alpha}%
}{\ln\frac{3\alpha}{2-4\alpha}},\frac{3\ln\varepsilon-\ln\frac{\alpha}{2}}%
{\ln\frac{3\alpha}{2\left(  1-\alpha\right)  }}\right\}  ,$}%
\end{equation}
such that%
\begin{equation}
\eta_{0}^{\prime}<\left(  \frac{3\alpha}{2\left(  1-\alpha\right)  }\right)
^{n_{b}}\frac{\alpha}{2}<\varepsilon^{3}%
\end{equation}%
\begin{equation}
\frac{\eta_{0}^{\prime}}{\delta_{0}^{\prime}}=\frac{\eta_{n_{b}}}%
{\delta_{n_{b}}}<\left(  \frac{3\alpha}{2-4\alpha}\right)  ^{n_{b}}%
\frac{\alpha}{1-3\alpha}<\varepsilon
\end{equation}
and%
\begin{equation}
\delta_{0}^{\prime}>\left(  \frac{1-2\alpha}{1-\alpha}\right)  ^{n_{b}}%
\frac{1-3\alpha}{2}%
\end{equation}

Then we choose for $n_{p}$
\begin{equation}
\fbox{$\frac{2\ln\varepsilon}{\ln\left(  1-2\delta_{0}^{\prime}\right)
}>n_{p}>\frac{\ln\varepsilon}{\ln\left(  1-2\lambda_{0}^{\prime}\left(
1-2\zeta\right)  \right)  }.$}%
\end{equation}
Such $n_{p}$ always exists, for $\frac{2\ln\varepsilon}{\ln\left(
1-2\delta_{0}^{\prime}\right)  }\approx\frac{2\ln\varepsilon}{-2\delta
_{0}^{\prime}}>\frac{\ln\varepsilon}{-2\delta_{0}^{\prime}\sqrt{\left(
1+\zeta\right)  \left(  1-2\zeta\right)  }}\approx\frac{\ln\varepsilon}%
{\ln\left(  1-2\lambda_{0}^{\prime}\left(  1-2\zeta\right)  \right)  }$. Thus,
we have
\begin{equation}
\eta_{n_{p}}^{\prime}<\left(  1+2\delta_{0}^{\prime}\right)  ^{n_{p}}\eta
_{0}^{\prime}<\left(  1-2\delta_{0}^{\prime}\right)  ^{-n_{p}}\eta_{0}%
^{\prime}<\varepsilon^{-2}\eta_{0}^{\prime}<\delta_{0}^{\prime}%
\end{equation}
and%
\begin{equation}
1/2-\lambda_{n_{p}}^{\prime}<\left(  1-2\delta_{0}^{\prime}\left(
1-2\zeta\right)  \right)  ^{n_{p}}\left(  1/2-\lambda_{0}^{\prime}\right)
<\varepsilon/2
\end{equation}

We now verify the two required relations which are underlined in the previous
discussion. The relation \underline{$\frac{1}{2}>\delta_{n}^{\prime}%
>\delta_{0}^{\prime}>\eta_{n}^{\prime}>\eta_{0}^{\prime}$} is satisfied for
all $0<n\leq n_{p}$. And the relation \underline{$\delta_{0}^{\prime}%
\eta_{n_{p}}^{\prime}+\eta_{0}^{\prime}<\delta_{0}^{\prime}\zeta$} is also
implied, since $\eta_{n_{p}}^{\prime}+\eta_{0}^{\prime}/\delta_{0}^{\prime
}<\varepsilon^{-2}\eta_{0}^{\prime}+\varepsilon<$\underline{$2\varepsilon
\equiv:\zeta$}.

Finally, the achievable fidelity for the above choice of $n_{b}$ and $n_{p}$
is
\begin{align}
1-F_{n_{p},n_{b}}  &  =1/2-\delta_{n_{p}}^{\prime}=1/2-\left(  \frac{1-2\zeta
}{1+\zeta}\right)  ^{1/2}\lambda_{n_{p}}^{\prime}\nonumber\\
&  <1/2-\left(  \frac{1-2\zeta}{1+\zeta}\right)  ^{1/2}\left(  1/2-\varepsilon
\right)  <4\varepsilon.
\end{align}
The bit-phase two-level pumping thus allows us to approach $F=1$ with
arbitrary good precision.

\section{Markov Chain Model for Two-Level Pumping
\label{Markov Chain Model for Two-Level Pumping}}

Here we present the Markov chain model for two-level entanglement pumping.

The state transition diagram for two-level entanglement pumping is shown in
Fig.~\ref{fig:MarkovChain2}. We assume that the required pumping steps are
$n_{b}$ and $n_{p}$ for the two levels, respectively. Since two entangled
pairs are stored, we need two labels to track the intermediate state for
two-level entanglement pumping.

Here we use "$0,0$" to denote the initial state with no Bell pairs, "$0,j+1$"
for the state with one purified pair surviving $j$ steps of pumping at the
first level, "$k+1,j+1$" for the state with one purified pair surviving $k$
steps of pumping at the second level and one purified pair surviving $j$ steps
of pumping at the first level, "$k+1,\ast$" for the state with one purified
pair surviving $k$ steps of pumping at the second level and one purified pair
surviving $n_{b}$ steps of pumping at the first level, and "$\ast,0$" for the
final state with one purified pair surviving $n_{p}$ steps of pumping at the
second level.

For the first level pumping, the (success) transition probability from state
"$k,j$" to state "$k,j+1$" is $q_{j}$, while the (failure) transition
probability from state "$k,j$" to state "$k,0$" is $1-q_{j}$, for
$j=0,1,\cdots,n_{b}$. For the second level pumping, the (success) transition
probability from state "$k,\ast$" to state "$k+1,0$" is $Q_{k}$, while the
(failure) transition probability from state "$k,\ast$" to state "$0,0$" is
$1-Q_{k}$, for $k=0,1,\cdots,n_{p}$. Here the transition probabilities
$\left\{  q_{j}\right\}  $ and $\left\{  Q_{k}\right\}  $ can be calculated
according to the density matrices of the intermediate purified Bell pairs
\cite{Dur99}. The final state is self-trapped, and goes back to itself with
unit probability. Altogether there are $\left(  n_{b}+2\right)  \left(
n_{p}+1\right)  +1$ states.

In order to fulfill the requirement that each transition attempt consumes one
unpurified Bell pair, we need to contract the states of "$k,\ast$" and
"$k+1,0$" into one state, since this transition does not consume any
unpurified Bell pair. After the contraction, there are $\left(  n_{b}%
+1\right)  \left(  n_{p}+1\right)  +1$ states remaining.%

\begin{widetext}%

Therefore, we may use a (column) vector $\vec{P}$ with $\left(  n_{b}%
+1\right)  \left(  n_{p}+1\right)  +1$ elements to characterize the
probability distribution among all $\left(  n_{b}+1\right)  \left(
n_{p}+1\right)  +1$ states. From the $t$-th attempt to the $\left(
t+1\right)  $th attempt, the probability vector evolves from $\vec{P}\left(
t\right)  $ to $\vec{P}\left(  t+1\right)  $ according to the following rule%
\begin{equation}
\vec{P}\left(  t+1\right)  =\mathbf{M}~\vec{P}\left(  t\right)  ,
\end{equation}
and the transition matrix is%
\begin{equation}
\mathbf{M}=\left(
\begin{array}
[c]{ccccc}%
\mathbf{M}_{1}+q_{n_{b}}\left(  1-Q_{0}\right)  \mathbf{N}_{1} & q_{n_{b}%
}\left(  1-Q_{1}\right)  \mathbf{N}_{1} & \cdots & q_{n_{b}}\left(
1-Q_{2}\right)  \mathbf{N}_{1} & 0\\
q_{n_{b}}Q_{0}\mathbf{N}_{1} & \mathbf{M}_{1} &  &  & \\
& q_{n_{b}}Q_{1}\mathbf{N}_{1} & \cdots &  & \\
&  & \cdots & \mathbf{M}_{1} & \\
0 &  &  & q_{n_{b}}Q_{n_{p}}\mathbf{N}_{1} & 1
\end{array}
\right)  , \label{eq:TranMatrix2}%
\end{equation}
with sub-matrices:%
\begin{equation}
\mathbf{M}_{1}=\left(
\begin{array}
[c]{ccccc}%
0 & 1-q_{1} & \cdots & 1-q_{n_{b}} & 0\\
1 & 0 &  &  & \\
& q_{1} & \cdots &  & \\
&  & \cdots & 0 & \\
&  &  & q_{n_{b}} & 0
\end{array}
\right)  _{\left(  n_{b}+1\right)  \times\left(  n_{b}+1\right)  },
\end{equation}
and
\begin{equation}
\mathbf{N}_{1}=\left(
\begin{array}
[c]{cccc}%
0 & \cdots & 0 & 1\\
&  & 0 & 0\\
& \cdots &  & \cdots\\
0 &  &  & 0
\end{array}
\right)  _{\left(  n_{b}+1\right)  \times\left(  n_{b}+1\right)  }.
\end{equation}
%

\end{widetext}%

\section{Fault-Tolerant Preparation of $2^{n}$-Qubit GHZ State\label{GHZ}}

We consider fault-tolerant preparation of a $2^{n}$-qubit GHZ state with
$n\geq3$. We label $2^{n}$ registers by $0,1,\cdots,2^{n}-1$. The GHZ state
can be prepared in just $3$ (clock cycles): in the first clock cycle, we
perform PBM for register pairs $\left(  2m,2m+1\right)  $ with integer $m$; in
the second clock cycle, we perform PBM for pairs $\left(  2m,2m+2\right)  $
and $\left(  2m+1,2m+3\right)  $; in the last clock cycle, \ we perform PBM
for pairs $\left(  8m+3,8m+5\right)  $, $\left(  8m+4,8m+6\right)  $, $\left(
16m^{\prime}+7,16m^{\prime}+9\right)  $, $\left(  16m^{\prime}+8,16m^{\prime
}+10\right)  $, $\cdots$, $\left(  2^{n-1}+2^{n-2}-1,2^{n-1}+2^{n-2}+1\right)
$, and $\left(  2^{n-1}+2^{n-2},2^{n-1}+2^{n-2}+2\right)  \,$\ in parallel.

In order to prepare the specific GHZ\ state $\left\vert 00\cdots0\right\rangle
+\left\vert 11\cdots1\right\rangle $, we still need to perform bit-flip
operations for individual registers, which are determined by the measurement
outcomes for all PBMs. Suppose the error probability for each PBM is $p$.
Since the redundancy checks of the PBMs impose consistency requirements for
measurement outcomes (error detection), the probability for undetected errors
in measurement outcomes has been suppressed to $O\left(  p^{2}\right)  $ for
each PBM. To the leading order of $p$, we only need to consider the phase
errors from PBM that are not detected by the redundancy check. Thus, the total
error probability scales as $2^{n-1}3p$, and the error probability for each
register is only approximately $3p/2$. Therefore, we have demonstrated a
PBM-based scheme to prepare the GHZ state fault-tolerantly, which is efficient
in both time and physical-resources.

\bibliographystyle{apsrev}
\bibliography{Ref4DistributedQC}

\end{document}